\documentclass[pdflatex,sn-mathphys-num]{sn-jnl}

\usepackage{graphicx}%
\usepackage{multirow}%
\usepackage{amsmath,amssymb,amsfonts}%
\usepackage{amsthm}%
\usepackage{mathrsfs}%
\usepackage[title]{appendix}%
\usepackage{xcolor}%
\usepackage{textcomp}%
\usepackage{manyfoot}%
\usepackage{booktabs}%
\usepackage{algorithm}%
\usepackage{algorithmicx}%
\usepackage{algpseudocode}%
\usepackage{listings}%

\theoremstyle{thmstyleone}%
\theoremstyle{thmstyletwo}%

\theoremstyle{thmstylethree}%

\begin{document}

\title[Article Title]{Solar phased arrays-based wireless power transfer for commercial airlines can reduce energy costs and carbon emissions in the United States}


\author[1]{\fnm{Tianyi} \sur{Wang}}\email{bonny.wang@utexas.edu}
\equalcont{These authors contributed equally to this work.}

\author[2]{\fnm{Yiming} \sur{Xu}}\email{yiming.xu@utexas.edu}
\equalcont{These authors contributed equally to this work.}

\author[1]{\fnm{Jiseop} \sur{Byeon}}\email{jsbyeon@utexas.edu}

\author[2]{\fnm{Junfeng} \sur{Jiao}}\email{jjiao@austin.utexas.edu}

\author[1]{\fnm{Javad} \sur{Mohammadi}}\email{javadm@utexas.edu}

\author[1]{\fnm{Kara} \sur{Kockelman}}\email{kkockelm@mail.utexas.edu}

\author*[1]{\fnm{Christian} \sur{Claudel}}\email{christian.claudel@utexas.edu}

\author[3]{\fnm{Alexandre} \sur{Bayen}}\email{bayen@berkeley.edu}

\affil*[1]{\orgdiv{Department of Civil, Architectural, and Environmental Engineering}, \orgname{The University of Texas at Austin}, \orgaddress{\city{Austin}, \postcode{78712}, \state{TX}, \country{USA}}}

\affil[2]{\orgdiv{School of Architecture}, \orgname{The University of Texas at Austin}, \orgaddress{\city{Austin}, \postcode{78712}, \state{TX}, \country{USA}}}

\affil[3]{\orgdiv{Department of Electrical Engineering and Computer Sciences}, \orgname{University of California}, \orgaddress{\city{Berkeley}, \postcode{94720}, \state{CA}, \country{USA}}}



\abstract{
Decarbonizing aviation remains challenging because energy-dense jet fuels dominate beyond short-range operations, while batteries impose severe range and payload penalties. 
Here we evaluate a new infrastructure pathway in which utility-scale solar farms equipped with solar phased arrays wirelessly beam microwave power to hybrid-electric aircraft during cruise. 
Integrating 143,152 U.S. flight trajectories, 5,712 solar farms and wireless power transfer models, we quantify the spatial, temporal, and operational potential of this concept at continental scale. 
We find that benefits are highly concentrated in solar-rich, traffic-dense states and are dominated by short- and medium-range flights, accounting for nearly all delivered energy and cost savings. 
Schedule optimization and higher cruise altitudes further increase value by improving alignment between aircraft demand and beaming availability. 
Market penetration analysis reveals non-linear scaling between solar farm and flight adoption.
These results show that wireless power beaming is best understood as a corridor-specific strategy complementing other aviation decarbonization pathways.
}

\maketitle

International climate targets call for net-zero carbon emissions from aviation by mid-century~\cite{allan2023intergovernmental}, yet the sector remains one of the most difficult to decarbonize~\cite{vardon2022realizing}. 
Commercial aviation accounts for approximately 2.5\% of global CO2 emissions and 3.5\% of effective radiative forcing, with demand projected to grow 3-4\% annually through 2050~\cite{dray2022cost,bergero2023pathways}. 
Although operational improvements can moderate fuel consumption~\cite{kosir2019improvement}, deep emissions reductions require fundamentally different energy supply pathways~\cite{male2021us}. 
Sustainable aviation fuels remain constrained by feedstock availability and high production costs~\cite{watson2024sustainable}, hydrogen faces volumetric energy density limitations that complicate airframe integration~\cite{yusaf2024sustainable}, and fully electric aircraft are restricted by battery gravimetric energy density to short-range missions below roughly 500 km~\cite{viswanathan2019potential,schafer2019technological,sampson2024operation}. 
For medium- and long-range operations, hybrid-electric aircraft combining conventional energy carriers with electric propulsion represent a pragmatic intermediate step~\cite{wheeler2021electric,rendon2021aircraft,sayed2021review,viswanathan2022challenges}, but no single low-carbon fuel or onboard storage technology presently offers a scalable, system-level solution for commercial aviation.


Meanwhile, the rapid expansion of utility-scale solar photovoltaics (PV) has fundamentally reshaped the economics of electricity generation~\cite{mehedi2022life}. 
Solar PV is now among the least expensive sources of electricity in many regions~\cite{aleksandra2024role}, and pathways in which solar supplies nearly half of US electricity would require deployment across roughly 0.5\% of the national land area~\cite{prasanna2021storage,heath2022environmental}.
In high-penetration scenarios, solar generation already produces substantial midday surpluses that must be curtailed, exported or stored~\cite{brunet2022will,maka2024pathway}, underscoring a growing mismatch between when and where renewable electricity is produced and where it is most valuable. 
This spatial and temporal abundance of low-cost solar electricity motivates new forms of sector coupling that extend beyond conventional electrification.

\begin{figure}[htbp!]
\centering
\includegraphics[width=\textwidth]{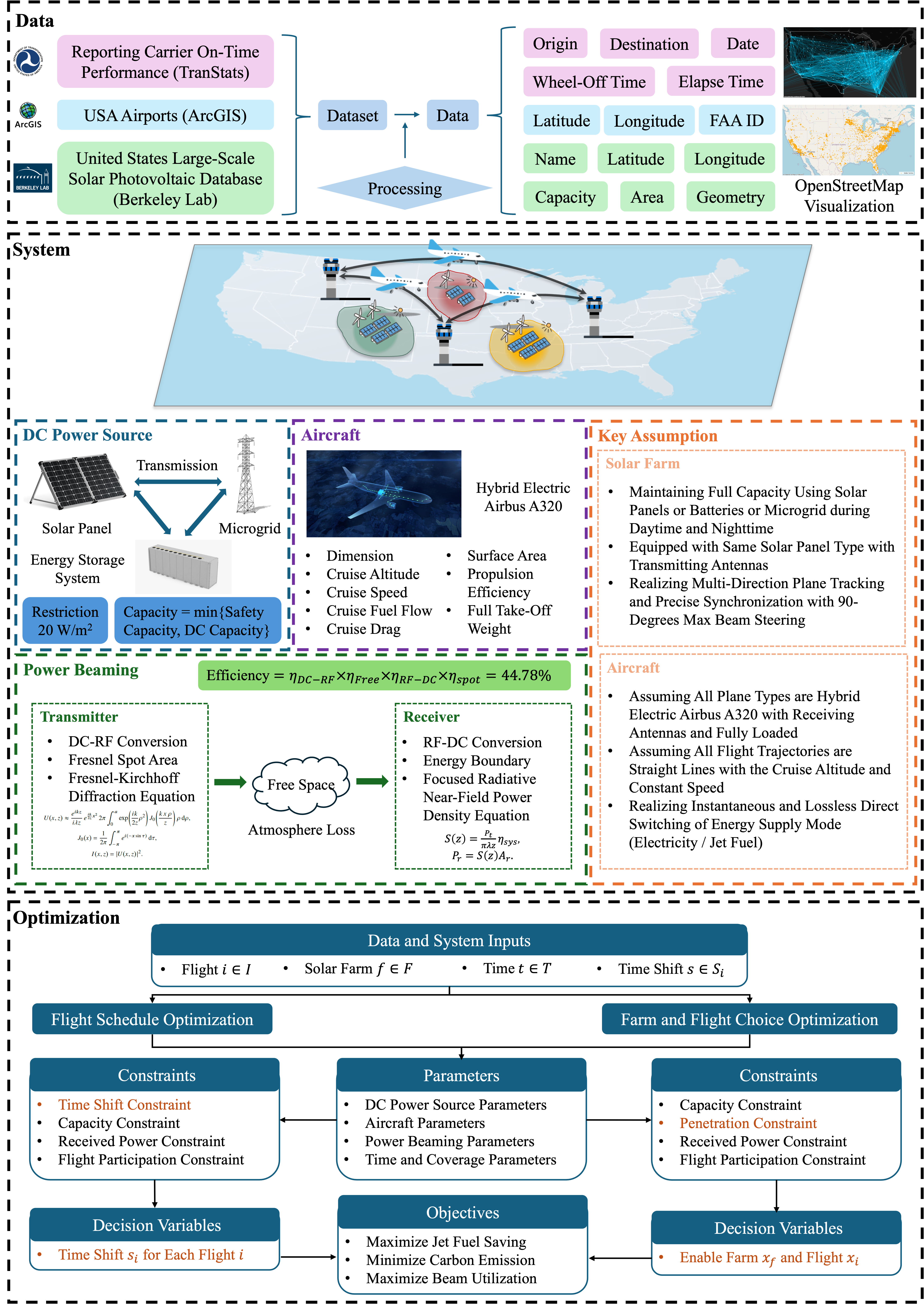}
\caption{Overview of methodology and data sources.
Top: Data inputs from three sources: the Reporting Carrier On-Time Performance dataset (TranStats) providing flight-level operational records (origin, destination, date, wheel-off time, elapse time); the USA Airports dataset (ArcGIS) providing airport coordinates and FAA identifiers; and the United States Large-Scale Solar Photovoltaic Database (Berkeley Lab) providing solar farm name, location, capacity, area and site geometry, are integrated through a geospatial processing pipeline. 
Middle: System model comprising the DC power source (solar panels with microgrid and energy storage, subject to a ground-level power density restriction of 20 $W/m^{2}$ and an effective capacity defined as the minimum of safety capacity and DC capacity), the hybrid-electric Airbus A320 aircraft with onboard receiving antenna, and the microwave power beaming subsystem (transmitter, free-space propagation and receiver) with an end-to-end efficiency of 44.78\%. 
Key assumptions for the solar farm and for the aircraft are summarized. 
Bottom: Optimization framework showing two complementary problems: flight schedule optimization and farm-and-flight choice optimization, sharing common parameters (DC source, aircraft, power beaming, time and coverage) and objectives (maximize fuel saving, minimize carbon emissions, maximize beam utilization).
}
\label{Picture1}
\end{figure}

One such pathway is wireless power transfer (WPT), in which electrical power is converted into microwave radiation, transmitted through free space and reconverted to electricity at a receiver~\cite{rodenbeck2021microwave}. 
The microwave frequency band is particularly suitable because the technologies are mature and cost-effective, antenna dimensions remain practical at utility scale, and atmospheric losses are low~\cite{shinohara2013beam}. 
Microwave power transmission was first demonstrated for a helicopter by Brown in 1964~\cite{brown2007experiments} and has since been validated at small scales for unmanned aerial vehicles~\cite{mahbub2024far}, electric vertical take-off and landing aircraft~\cite{wang2025large}, conventional tube-and-wing configurations~\cite{orndorff2023gradient}, high altitude platforms~\cite{zhao2020ka}, etc. 
Recent work proposed solar phased arrays (SoPhAr), i.e., solar PV farms equipped with embedded, phase-controlled emitters, to form and steer focused beams of microwave energy toward aircraft in flight~\cite{claudel2024sophar}. 
Because the effective aperture of a utility-scale solar farm can extend over hundreds of meters to kilometers, such systems can operate in the radiative near field, enabling controlled beam shaping and power delivery over tens of kilometers with manageable power densities~\cite{yu2018design}. 
Despite this physical promise, the system-level implications of coupling aviation energy demand to ground-based solar infrastructure via energy beaming remain largely unexplored. 
Existing studies have focused on the physics of beam formation~\cite{liu2024multifunctional,saha2024novel,silva2025multibeam} and the design of antennas~\cite{patwary20244,xia2024adaptive,roy2025wireless}. 
No integrated analysis has yet examined how a SoPhAr-based WPT system would interact with real-world flight networks, the spatial distribution of solar farms, electricity market conditions, and the operational constraints of continental-scale air traffic. 


To address these knowledge gaps, this study develops a quantitative, network-scale assessment of SoPhAr-based energy beaming for commercial aviation across the United States (Fig. \ref{Picture1}). 
Combining 143,152 flight trajectories recorded over one week (7th–13th April, 2025) with geospatial data for 5,712 utility-scale solar farms and physics-based WPT system models, we evaluate achievable coverage, power transfer potential and operational impacts across a continental-scale air traffic network. 
We further quantify how cruise altitude variation, optimized flight scheduling, and market penetration jointly influence energy costs and carbon emissions. 
By explicitly linking solar farm deployment to flight adoption, our analysis reframes energy beaming from a component-level concept into a coupled infrastructure problem, and quantifies its potential role in a deeply decarbonized aviation system.

\begin{figure}[htbp!]
\centering
\includegraphics[width=\textwidth]{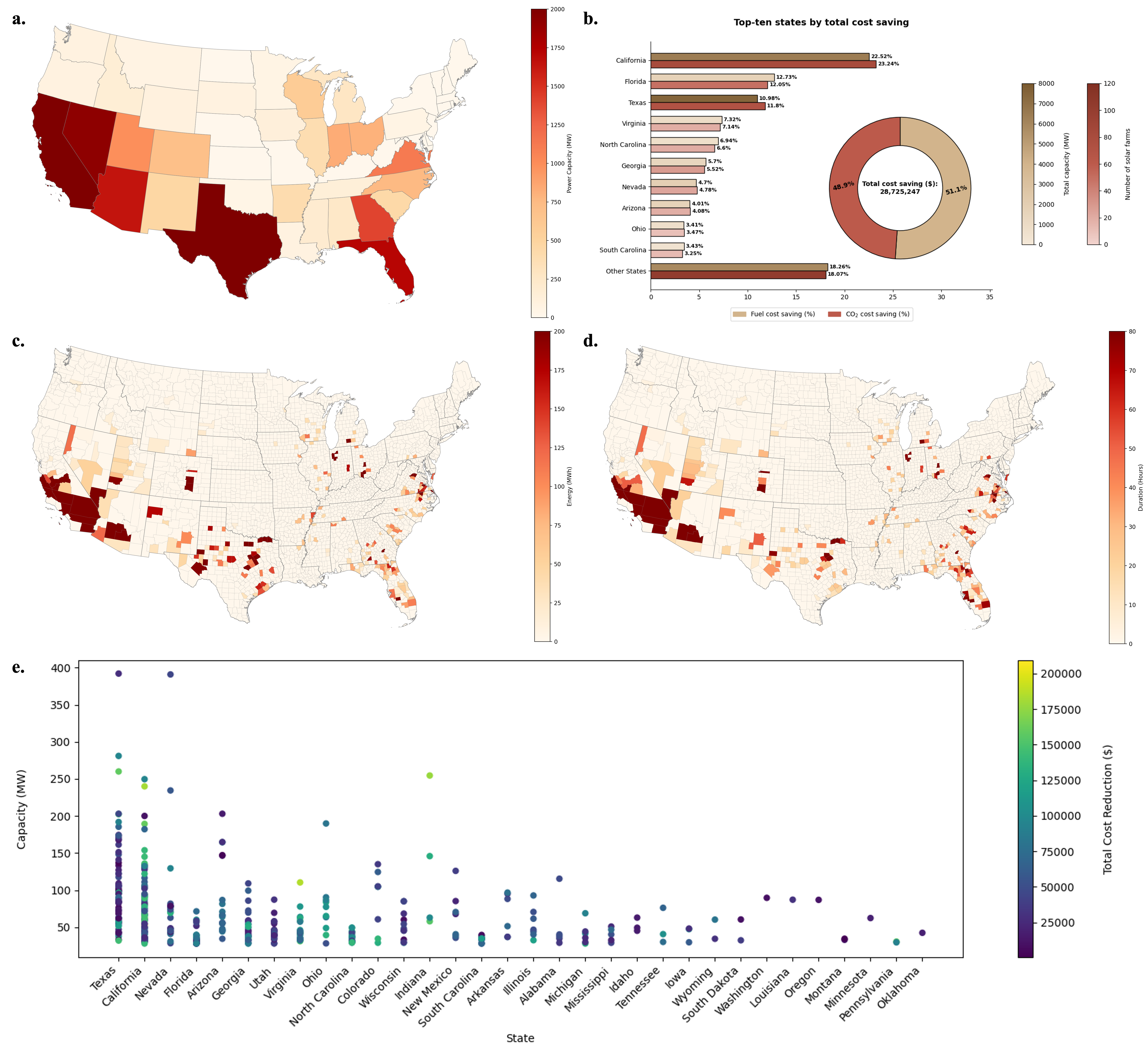}
\caption{Solar farm-level analysis of capacity, energy, duration and economic benefits in the SoPhAr-enabled network.
a, State-level distribution of total solar farm safety capacity (MW).
b, Top ten states ranked by total cost saving (\$), decomposed into fuel cost saving (\$) (brown) and CO2 cost saving (\$) (red). The donut chart summarizes the national split between fuel- and CO2-related components; and color bars indicate total capacity (MW) and number of solar farms in each state.
c, County-level spatial distribution of total transferred energy (MWh).
d, County-level spatial distribution of total effective beaming duration (h).
e, Relationship between individual solar farm capacity (MW) and total cost reduction (\$), colored by cost reduction magnitude.}
\label{Picture2}
\end{figure}

\section{Results}\label{sec2}



\subsection{Beaming value concentrates in solar-rich states}


The economic benefits of SoPhAr-enabled energy beaming are highly unevenly distributed across the United States. 
Of the 5,712 solar farms in the dataset, 437 meet the qualification criteria for participation in the beaming network, representing 29.35 GW of total safety capacity. 
Over the one-week analysis period, these farms yield an estimated \$14.68M in fuel cost reduction and 73.92M kg of avoided CO2 emissions.


This potential concentrates in the solar-rich South and West (Fig. \ref{Picture2}a). 
California provides the single largest contribution, with 6.43 GW of capacity distributed across 82 farms and an estimated \$6.57M in total cost saving, corresponding to 22.5\% of national fuel cost savings and 23.2\% of CO2 cost savings (Fig. \ref{Picture2}b). 
Florida follows with 1.70 GW across 49 farms and \$3.56M in total savings (12.7\% and 12.1\% of fuel and CO2 savings, respectively). 
Texas, despite ranking first in installed capacity at 7.53 GW across 71 farms, contributes a smaller share of total cost saving (11.0\% fuel, 11.8\% CO2). 
This mismatch between capacity and economic ranking indicates that installed capacity alone does not determine value. 
Rather, economic performance depends on how effectively that capacity is positioned relative to high-density flight corridors and how long aircraft remain within effective beaming range. 
Notably, at both national and state level, fuel and CO2 cost savings track closely, indicating that solar farm deployment decisions optimizing direct operating cost savings are also likely to capture a substantial fraction of the climate-related economic benefit.


County-level maps of supplied energy (Fig. \ref{Picture2}c) and supported flight duration (Fig. \ref{Picture2}d) reveal that performance clusters in a limited number of local areas rather than distributing uniformly within states. 
These two metrics should be interpreted together: energy captures the magnitude of transferred power, whereas duration reflects the persistence of operational opportunity. 
States that score well on both dimensions tend to dominate total cost reduction.
The farm-level scatter plot (Fig. \ref{Picture2}e) reinforces this interpretation. 
Although larger farms tend to unlock greater savings, several medium-capacity farms produce savings comparable to or exceeding those of larger installations. 
This dispersion demonstrates that the most valuable solar farms are those embedded in locations where local solar capacity and aircraft traffic geometry coincide most favorably, a deployment problem that is fundamentally network-dependent rather than purely capacity-driven.

\begin{figure}[htbp!]
\centering
\includegraphics[width=\textwidth]{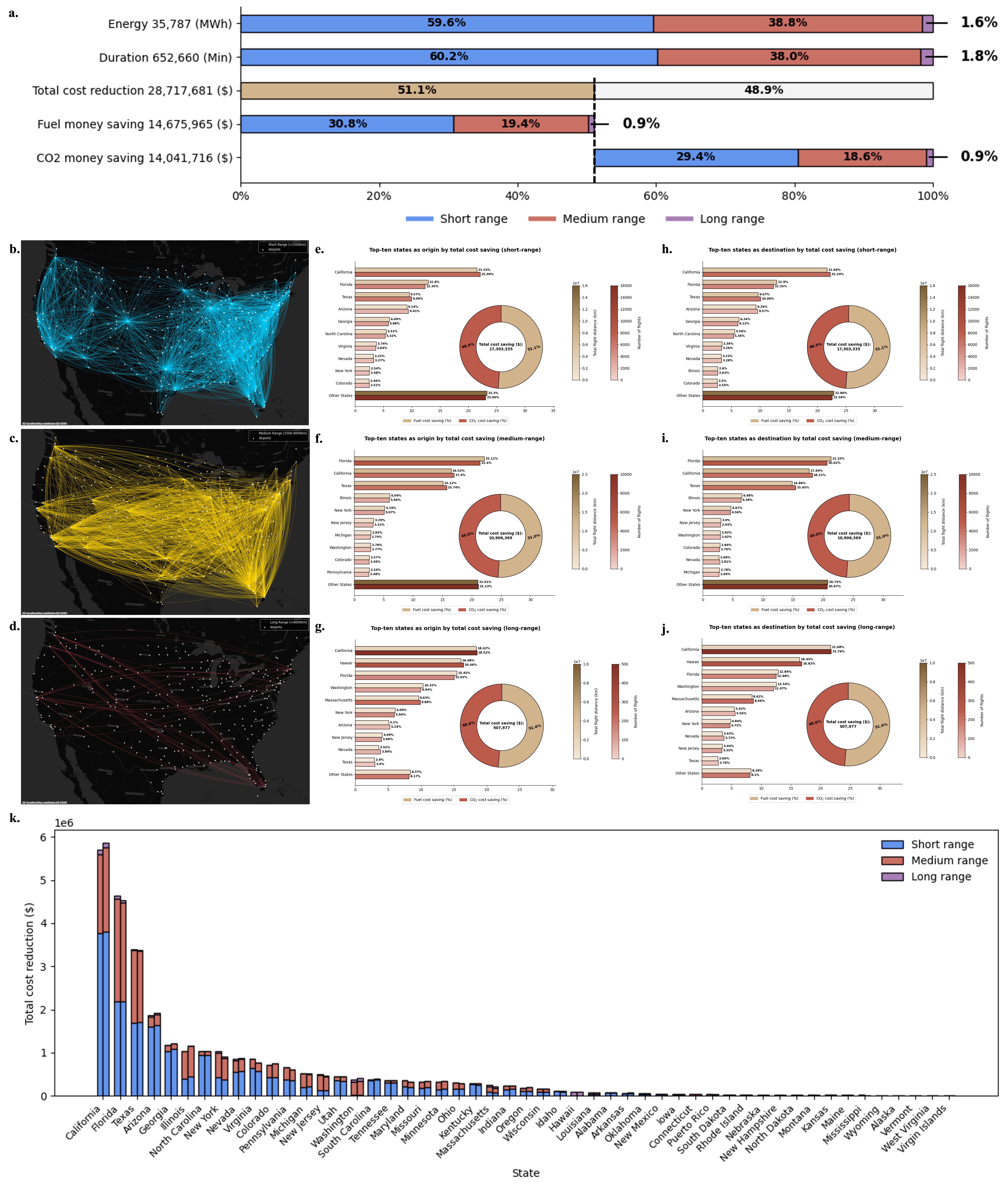}
\caption{Flight-level analysis of energy, duration and economic benefits across different flight range classes.
a, Range-class decomposition of system-wide outcomes, showing shares of transferred energy (MWh), supported duration (min), fuel cost saving (\$), CO2 cost saving (\$) and total cost reduction (\$) attributable to short-range (0–1,500 km), medium-range (1,500–4,000 km) and long-range ($>$4,000 km) flights.
b–d, Spatial distributions of eligible flight connections for short-range (b), medium-range (c) and long-range (d) flights.
e–g, Top ten origin states ranked by total cost saving for short-range (e), medium-range (f), and long-range (g) flights; horizontal bars decompose fuel and CO2 cost savings, donut charts summarize the corresponding split between fuel- and CO2-related savings, and color bars indicate total flight distance (km) and number of flights.
h–j, Top ten destination states ranked by total cost saving for short-range (h), medium-range (i), and long-range (j) flights, with the same decomposition into fuel and CO2 cost saving.
k, State-level total cost reduction aggregated across all flights, decomposed by range class.}
\label{Picture3}
\end{figure}



\subsection{Short- and medium-range corridors drive economic value}


At the flight level, the value of SoPhAr-enabled WPT is overwhelmingly concentrated in short- and medium-range operations. 
Across all analyzed flights, the system delivers 35,787 MWh of beamed energy and 652,660 min of supported operation, yielding a total cost reduction of \$28.72M, of which \$14.68M arises from avoided fuel expenditure and \$14.04M from avoided CO2 costs.

Short-range flights (0–1,500 km) account for 59.6\% of delivered energy and 60.2\% of supported duration, while medium-range flights (1,500–4,000 km) contribute a further 38.8\% and 38.0\%, respectively (Fig. \ref{Picture3}a). 
Long-range flights ($>$4,000 km) represent only 1.6\% of supplied energy and 1.8\% of supported duration. 
The monetary pattern mirrors this concentration: short-range flights contribute 30.8\% of total fuel cost savings and 29.4\% of CO2 cost savings, medium-range flights add 19.4\% and 18.6\%, and long-range flights account for less than 1\% of each component. 
This imbalance indicates that the economic value of SoPhAr is driven by dense short- and medium-range corridors rather than by the relatively sparse set of long-range routes.


Route maps (Fig. \ref{Picture3}b–d) clarify this concentration. 
Short-range benefits distribute across a dense national mesh with especially strong activity in the eastern United States, while medium-range benefits show a broader coast-to-coast structure linking major population centers and hub airports. 
Long-range benefits are visibly sparse, with only a limited set of routes contributing meaningful savings. 
SoPhAr-based WPT is more valuable when flights intersect multiple beaming windows over solar-rich and traffic-dense regions.
Such opportunities arise frequently on short- and medium-range corridors, but much less often on long-range routes, where flight paths are more spatially dispersed and less consistently aligned with clusters of utility-scale solar farms.
State-level origin and destination rankings (Fig. \ref{Picture3}e–j) further reinforce this interpretation. 
For short-range flights, California is the dominant state, followed by Florida and Texas, with Arizona, Georgia and North Carolina forming a second tier. 
For medium-range flights, Florida becomes the leading contributor, ahead of California and Texas. 
For long-range flights, the ranking shifts markedly: Hawaii, Washington and Massachusetts rise in prominence, reflecting the geography of the limited long-distance corridors that still intersect economically valuable beaming opportunities. 
The close similarity between origin- and destination-based rankings suggests that SoPhAr benefits accrue along bidirectional corridor structures rather than being concentrated at one end of the trip. 
The stacked state bars (Fig. \ref{Picture3}k) show that states such as California, Florida and Texas derive value from multiple distance classes simultaneously, whereas many lower-ranked states derive nearly all benefit from a single range class, highlighting a much narrower operational role in the SoPhAr-enabled network.

\begin{figure}[htbp!]
\centering
\includegraphics[width=\textwidth]{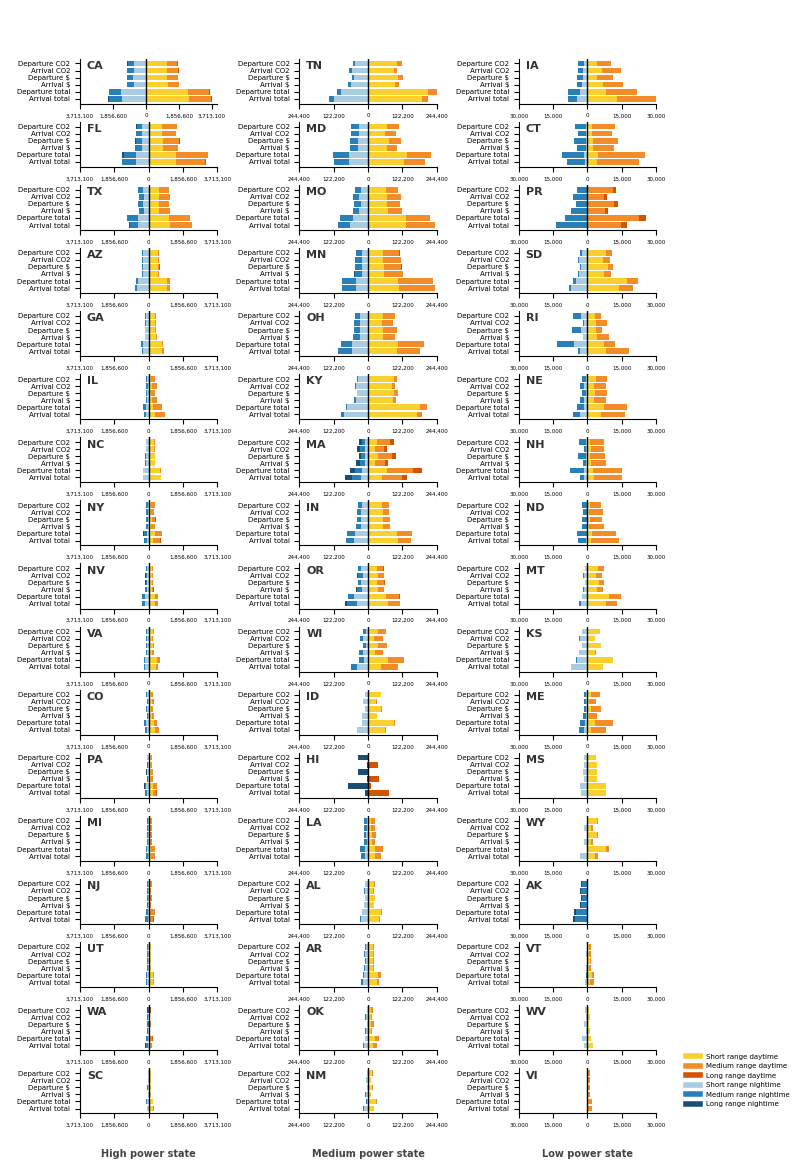}
\caption{Temporal analysis of economic benefits across high-, medium- and low-power states.
State-level bar charts compare nighttime (18:00–06:00) and daytime (06:00–18:00) contributions of departure- and arrival-based CO2 cost saving (\$), fuel cost saving (\$) and total cost reduction (\$) for short-range (0–1,500 km), medium-range (1,500–4,000 km) and long-range ($>$4,000 km) flights. 
States are grouped into high-power states (left column), medium-power states (middle column) and low-power states (right column) according to their overall solar power and beaming potential.}
\label{Picture4}
\end{figure}



\subsection{Daytime operations dominate system-wide value}

Temporal disaggregation reveals that daytime operations (06:00–18:00) dominate system-wide value across the majority of states and range classes (Fig. \ref{Picture4}), a result consistent with the direct availability of solar generation during daylight hours and the concentration of commercial flight departures and arrivals during daytime scheduling windows. 
This alignment between peak solar output and peak air traffic density is a structural advantage of SoPhAr-based WPT, which delivers the greatest value precisely when both supply and demand are at their highest.

Among high-power states, particularly California, Florida and Texas, the largest departure and arrival cost reductions, as well as the largest fuel and CO2 savings, are associated with daytime operating windows. 
Across nearly all states, short-range flights provide the largest contribution in both daytime and nighttime periods, with medium-range flights forming the second-largest component and long-range flights remaining comparatively small. 
This pattern mirrors the flight-level analysis, but the temporal view sharpens the imbalance: short-range operations dominate not only because they are numerous, but also because they are frequent enough throughout day and night to sustain significant value. 
The small distinction between origin- and destination-based rankings confirms that SoPhAr benefits accrue along bidirectional corridor structures rather than being concentrated at one end of the trip.
Medium-power states display a more varied temporal structure, though daytime savings generally remain dominant. 
Hawaii is a notable exception, showing a pronounced diurnal alternation: high departure value during nighttime and high arrival value during daytime, both driven exclusively by long-range flights, a pattern reflecting its geographic isolation and the timing of transcontinental connections. 
Low-power states reveal a different operational regime entirely, where total savings are typically concentrated in medium-range flights with limited short-range contributions and almost negligible long-range value. 

Nighttime savings are consistently smaller in magnitude, indicating that grid-connected or stored electricity at solar farms can extend value beyond daylight hours, but the core economic case for SoPhAr rests on the daytime operating window.
These findings suggest that the near-term deployment case for SoPhAr is strongest in states where daytime solar generation and dense daytime flight corridors coincide, a condition already met in California, Texas and Florida. 
Nighttime value, while additive, should be viewed as a secondary benefit that could grow over time as storage costs decline and grid-connected solar farms extend their dispatchable capacity into evening and overnight hours.



\begin{figure}[htbp!]
\centering
\includegraphics[width=\textwidth]{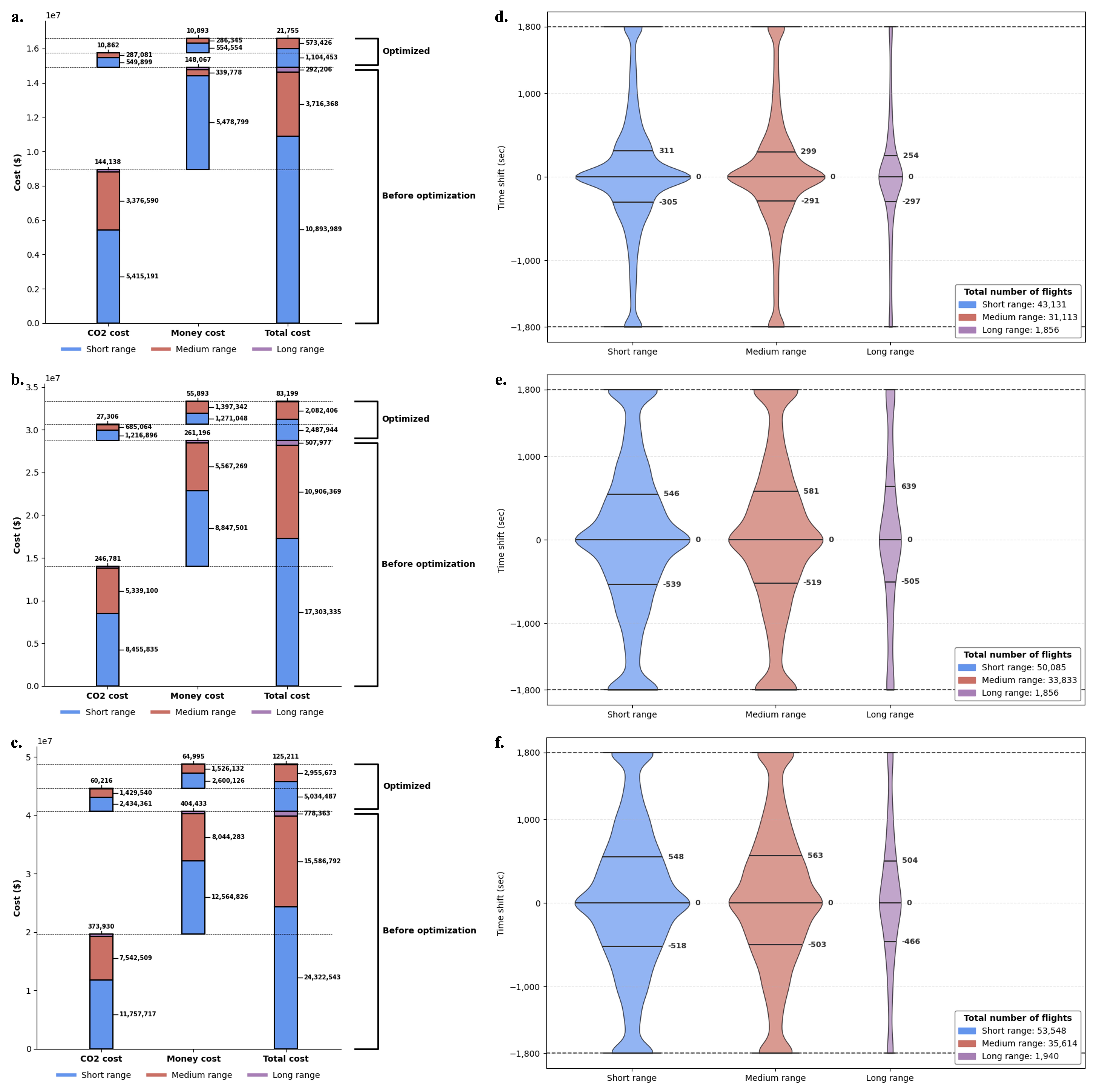}
\caption{Cruise altitude and schedule optimization effects across different flight range classes.
a-c, Stacked bar charts showing CO2 cost (\$), fuel cost (\$) and total cost reduction (\$) for short-range (0-1,500 km), medium-range (1,500-4,000 km) and long-range ($>$4,000 km) flights before (lower bars) and after (upper bars) schedule optimization at three cruise altitudes: 9,100 m (a), 12,100 m (b), and 15,100 m (c). 
d-f, Violin plots of optimized time shifts for short-range, medium-range and long-range flights at each altitude: 9,100 m (d), 12,100 m (e), and 15,100 m (f). 
Distributions are bounded by the allowable shift limits of ±1,800 s; and annotated quartile values indicate the spread of schedule adjustments within each range class. 
Insets report the total number of flights affected.}
\label{Picture5}
\end{figure}

\subsection{Cruise altitude and schedule optimization amplify benefits}


Varying cruise altitude and permitting limited flight schedule adjustment substantially alter the economic performance of SoPhAr-enabled energy beaming, but the gains are not uniform across distance classes (Fig. \ref{Picture5}). 
Across all evaluated conditions, short-range flights remain the dominant source of total benefit, medium-range flights provide the second-largest contribution, and long-range flights remain comparatively small even after optimization. 
At the standard cruise altitude of 12,100 m (Fig. \ref{Picture5}b), optimization increases total cost reduction from \$28.72M to \$33.37M, an improvement of \$4.65M, composed of an increase in CO2 cost saving from \$14.04M to \$15.97M and an increase in fuel cost saving from \$14.68M to \$17.40M. 
Short-range flight savings rise from \$17.30M to \$19.79M, medium-range from \$10.91M to \$12.99M and long-range from \$0.51M to \$0.59M. 
In relative terms, medium-range flights show the clearest proportional improvement, suggesting that schedule shifting particularly helps flights whose baseline alignment with beaming windows is good but not yet optimal.

The altitude comparison (Fig. \ref{Picture5}a–c) reveals a clear ordering: 15,100 m yields the largest total system value and the largest optimization gain, 12,100 m yields intermediate value, and 9,100 m yields the lowest. 
This indicates that higher cruise altitude expands the effective beaming footprint, and realizing the full benefit requires corresponding schedule coordination. 
The stronger optimization response at 15,100 m implies that altitude and timing should be treated as coupled design variables rather than as independent operational choices.

The schedule-shift distributions (Fig. \ref{Picture5}d–f) clarify how these gains are achieved. 
In all cases, the allowable optimization space is fully used, with shifts reaching the imposed bounds of ±1,800 s. 
At 9,100 m, interquartile ranges are concentrated around zero (approximately -305 s to +311 s for short-range flights). 
At 12,100 m, the distributions widen substantially (-539 s to +546 s for short-range flights), and a similar pattern appears at 15,100 m (-518 s to +548 s). 
In all cases, the median remains at zero, indicating that the optimization does not require systematic bias toward earlier or later departures but rather redistributes flights around their original schedules to better align with available beaming opportunities. 
The number of affected flights also increases with altitude, from 43,131 short-range flights at 9,100 m to 53,548 at 15,100 m, reinforcing the view that altitude expands the operational value of coordination rather than replacing it.







\subsection{Non-linear scaling with market penetration rates}



\begin{figure}[htbp!]
\centering
\includegraphics[width=\textwidth]{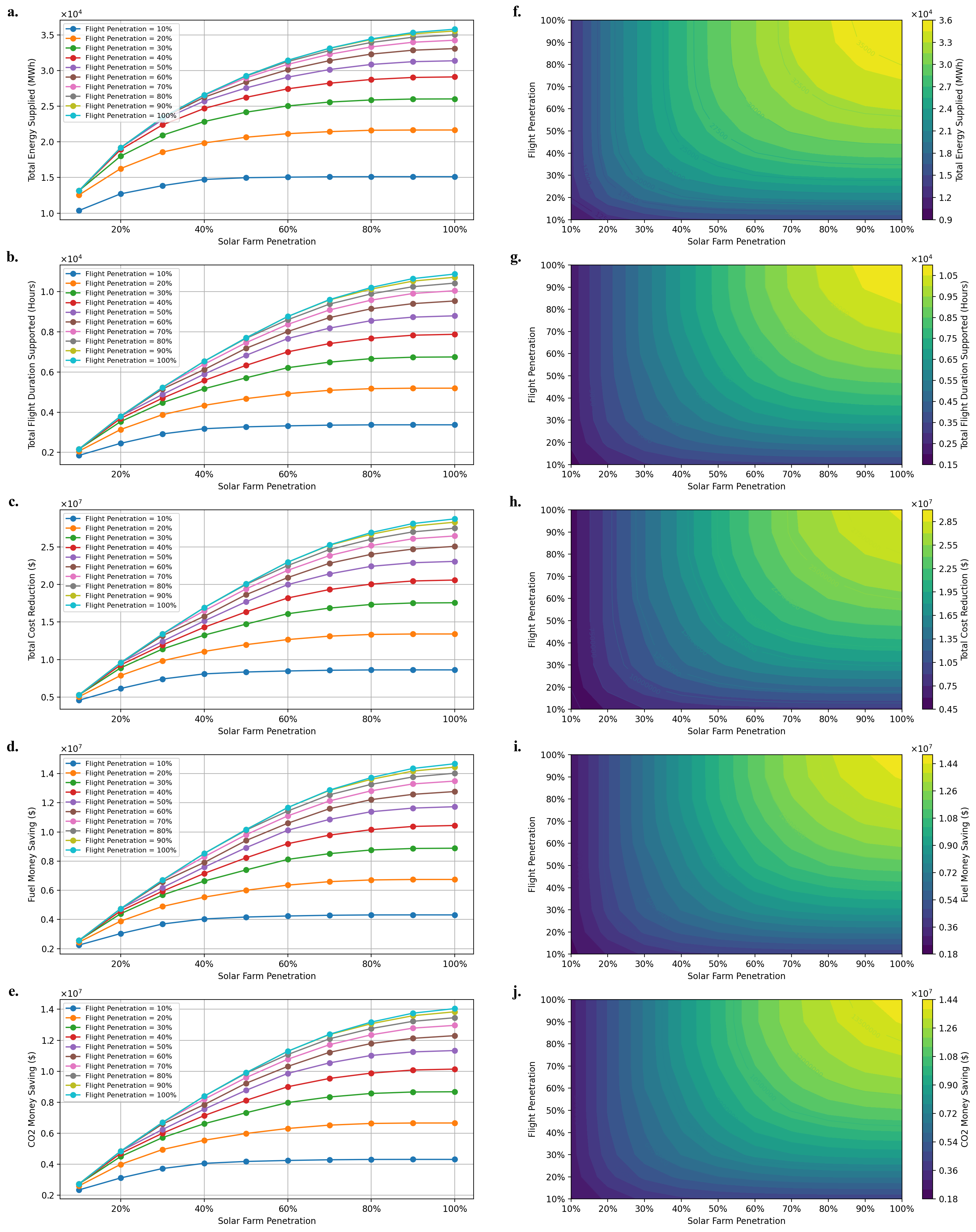}
\caption{Optimal system performance across different market penetration scenarios.  
a-e, Line charts detailing the isolated effect of solar farm market penetration rates (0-100\%) on five primary operational and economic metrics, with individual curves representing different flight market penetration rates. 
f-j, Corresponding two-dimensional contour plots illustrating the joint effects of solar farm and flight penetration on the same metrics. 
Evaluated metrics are: total energy supplied (MWh) (a,f); total flight duration supported (hours) (b,g); total cost saving (\$) (c,h); fuel cost saving (\$) (d,i); and CO2 cost saving (\$) (e,j).
Color gradients denote the magnitude of respective metrics.
}
\label{Picture6}
\end{figure}

The relationship between solar farm and flight market penetration rates reveal positive, non-linear scaling across all measured systemic metrics (Fig.~\ref{Picture6}). 
In general, increasing both solar farm penetration and flight penetration drives substantial gains in total energy supplied, supported flight duration, and overall economic savings, but with diminishing marginal returns (Fig.~\ref{Picture6}a-e). 
While initial increases in solar farm capacity (from 10\% to roughly 40\%) yield rapid expansions in system capabilities and cost reductions, these benefits distinctly plateau at higher penetration levels.
This saturation effect is dependent on flight penetration: at lower flight penetration rates (10–20\%), the system quickly reaches a bottleneck, indicating that excess solar generation remains underutilized without sufficient demand from flights. 
The contour plots (Fig.~\ref{Picture6}f-j) further emphasize this dependency, with the steepest gradients occurring in the lower-left quadrants, while the absolute maximum systemic benefits are localized to the upper right (100\% penetration for both variables). 
Ultimately, these results indicate that solar infrastructure investments must be strategically coupled with flight electrification demand to prevent overcapacity and maximize systemic benefits.

\begin{figure}[htbp!]
\centering
\includegraphics[width=\textwidth]{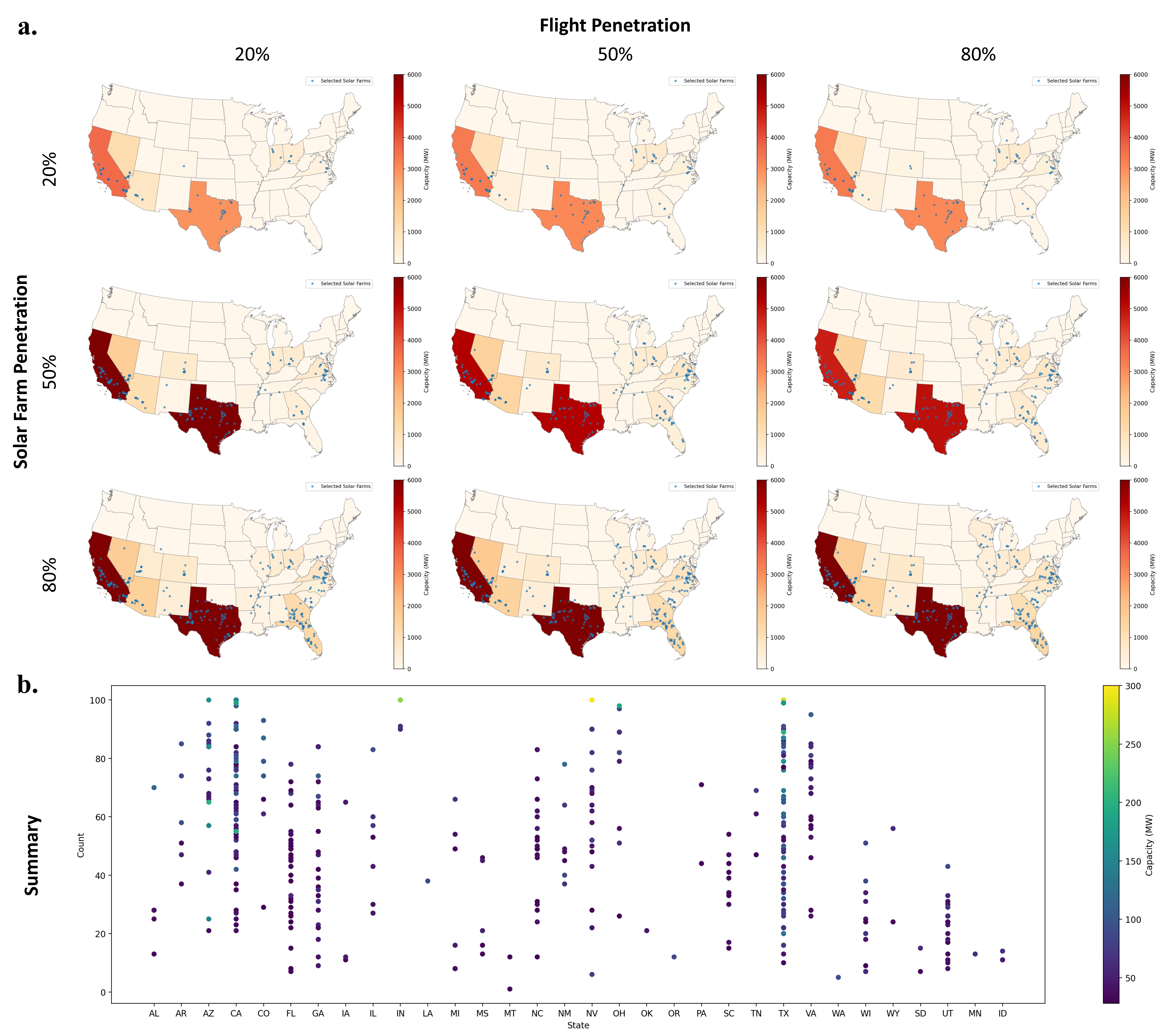}
\caption{Spatial patterns and robustness of optimized solar farm selection. 
a, Spatial distribution of selected solar farms (blue dots) across different scenarios of solar farm penetration (rows: 20\%, 50\%, 80\%) and flight penetration (columns: 20\%, 50\%, 80\%). 
The color of the states represents the total optimized solar capacity (MW). 
b, Scatter plot aggregates the frequency and capacity of individual solar farms selected across all 100 penetration scenarios (10\% intervals for both solar farm and flight penetration). 
Each dot represents one farm, clustered by state; vertical position indicates how many of the 100 scenarios selected that farm; and color corresponds to the capacity of that solar farm (MW).}
\label{Picture7}
\end{figure}

The optimization results reveal a highly concentrated and invariant spatial distribution for solar farms selection (Fig.~\ref{Picture7}a). 
Optimal capacity allocation is consistently concentrated in California, Texas, and the Southeast (notably Georgia and Florida), regardless of penetration rates. 
This stability suggests that underlying geographical advantages, such as localized solar irradiance and proximity to critical aviation demand networks, outweigh variable operational constraints.
A subset of solar farms is selected in nearly all 100 modeled penetration permutations (Fig.~\ref{Picture7}b), and this universal prioritization is not strictly dependent on solar farm capacity. 
Both massive utility-scale solar farms (capacity around 300 MW) and smaller arrays are consistently selected. 
This decoupling of selection frequency from facility capacity indicates that geographical location is the primary driver of operational utility. 
Consequently, these scenario-agnostic sites constitute optimal targets for early-stage infrastructural investment, as their systemic value remains resilient to uncertainties in future electrification adoption rates.

\begin{figure}[htbp!]
\centering
\includegraphics[width=\textwidth]{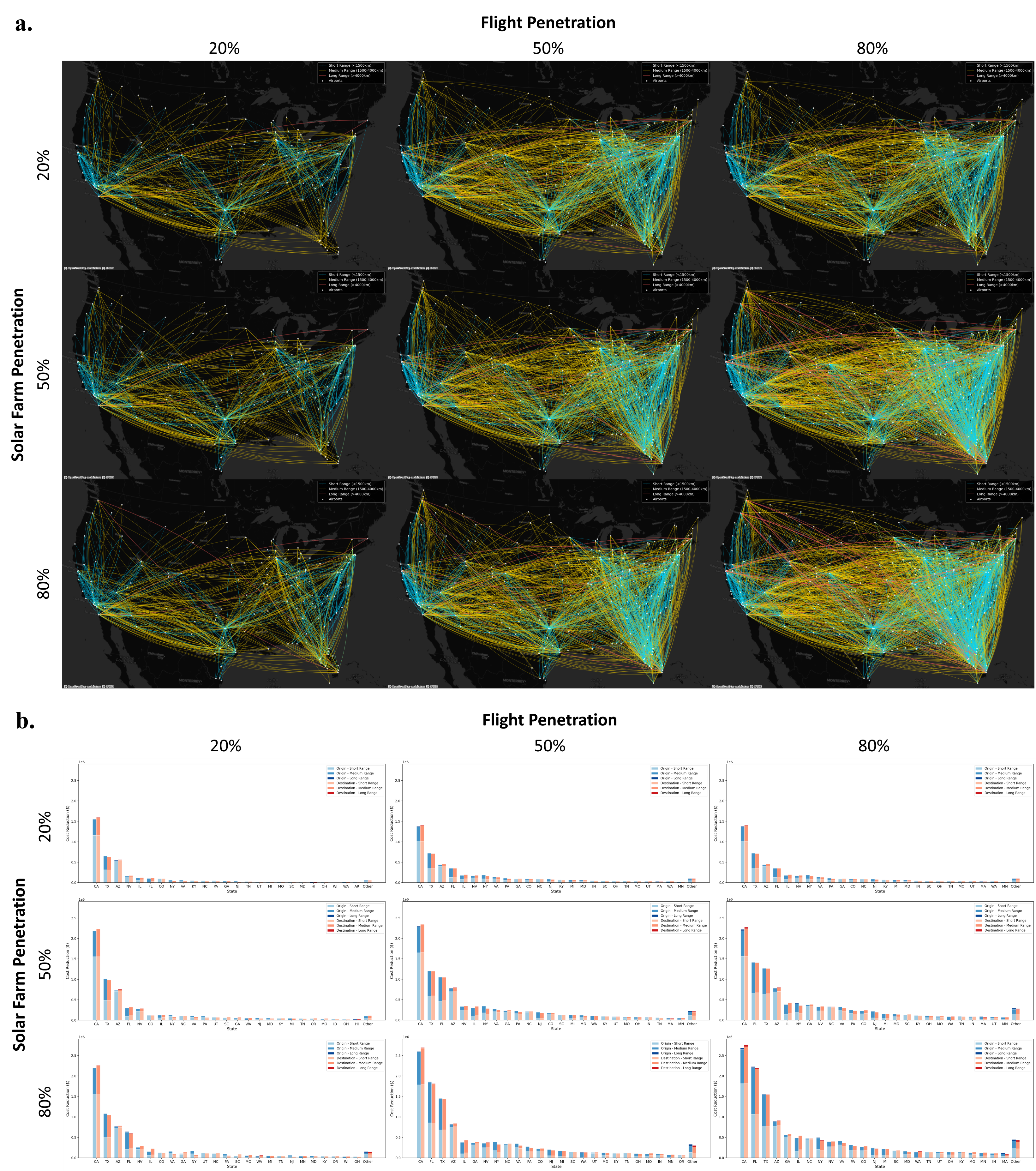}
\caption{Optimized flight networks and state-level cost reductions under different market penetration rates. 
a, Selected flights maximizing system-wide benefits across different scenarios of solar farm penetration (rows: 20\%, 50\%, 80\%) and flight penetration (columns: 20\%, 50\%, 80\%). 
Each line represents a flight connection colored flight range category: short-range (0-1,500 km, cyan), medium-range (1,500-4,000 km, yellow), and long-range ($>$4,000 km, magenta). 
Airports are shown as white points. 
b, State-level cost reductions aggregated by flight origin and destination states, with states ordered by total savings (\$). 
For each state, the left bar (blue shades) represents cost savings associated with flight origins, and the right bar (orange shades) represents savings associated with destinations. 
Bar segments are stacked by flight range category.}
\label{Picture8}
\end{figure}

Across all penetration scenarios, the optimal system-level savings are predominantly generated by dense, high-frequency corridors along the West Coast, the Eastern Seaboard, and major transcontinental gateways (Fig.~\ref{Picture8}a). 
The network of selected flights is dominated by short- and medium-range flights, while long-range flights account for only a minor share. 
State-level cost reductions are dominated by California, Texas, and Florida, irrespective of market penetration rates, with economic benefits distributed symmetrically between origins and destinations (Fig.~\ref{Picture8}b), underscoring the bidirectional nature of these high-value aviation corridors. 
These findings suggest that the most effective near-term pathway for aviation electrification is to prioritize short- and medium-range corridors centered on these high-capacity states.



\section{Discussion}\label{sec3}


This study demonstrates that the value of SoPhAr-based WPT is shaped not by beam physics or aircraft design, but by the spatial overlap between solar infrastructure and flight corridors, the temporal alignment between power availability and traffic demand, and the operational flexibility of hybrid-electric airlines. 
Benefits are highly concentrated: a relatively small number of solar-rich states, dense short- and medium-range corridors, and operationally favorable time windows account for most of the modeled energy delivery, cost savings and emissions reduction.
A credible deployment pathway focused on high-value regional clusters, especially in states such as California, Florida and Texas, where both solar infrastructure and flight activity are already concentrated. 
For policymakers, this points to corridor-based planning; for airlines, to targeted receiver integration on high-frequency routes; and for investors, to evaluating SoPhAr as a transport-linked energy asset whose return depends on corridor traffic, aircraft adoption and operational coordination.


The contribution of nighttime operations indicates that SoPhAr cannot be interpreted simply as a daylight solar application; its value extends through interacting with storage systems and the broader grid, positioning SoPhAr at the intersection of electricity markets, storage policy and air transport operations. 
Regulatory frameworks for interconnection, storage co-location, electricity pricing and dispatch flexibility may therefore be as important as improvements in beam efficiency or receiver design.
Operational optimization further increases system value: higher cruise altitude expands the effective beaming footprint, and modest schedule adjustments allow flights to align more effectively with available beaming windows, suggesting that altitude and scheduling should be treated as coupled design variables. 
The market penetration analysis reveals that system value scales non-linearly with both solar farm and flight adoption rates, exhibiting diminishing marginal returns.
This coupled scaling behavior underscores that SoPhAr should be planned as a coordinated infrastructure–flight roll-out rather than sequential supply-side deployment.


The analysis relies on assumptions regarding system efficiency, aircraft characteristics, safety constraints and market conditions and does not resolve practical issues surrounding certification, microwave exposure governance, or land-use acceptance. 
This study covers a single operating period in the US network.
Future work should test robustness under seasonal variability, alternative electricity market structures and international routes.


\section{Methods}\label{sec4}

\subsection{Overview}

Our methodology integrates three components: data, system, and optimization, to evaluate the network-scale performance of SoPhAr-based WPT for commercial aviation across the United States (Fig. \ref{Picture1}). 
We first assemble datasets that characterize the national flight network, airport locations and utility-scale solar farm infrastructure. 
These datasets are then coupled with physics-based models of the direct-current (DC) power source, hybrid-electric aircraft and microwave power beaming subsystem to compute the instantaneous power that each solar farm can deliver to each aircraft as a function of position, time and system constraints. 
Finally, we formulate two complementary mixed-integer linear programs (MILPs): flight schedule optimization and farm-and-flight choice optimization, to quantify the maximum achievable fuel cost savings, CO2 emission reductions and beam utilization under realistic operational constraints. 
We further classify flights into three distance categories: short-range (0-1,500 km), medium-range (1,500-4,000 km) and long-range ($>$4,000 km), and disaggregate results by daytime (06:00-18:00) and nighttime (18:00-06:00) operating windows to reveal the temporal and operational structure of beaming value.
Each component in the proposed method can be updated independently as technology matures or data becomes available, and the framework can be extended to other geographies or aircraft types.

\subsection{Data}

\subsubsection{Flight data}

We use the Reporting Carrier On-Time Performance dataset \cite{flightdata} from the Bureau of Transportation Statistics (TranStats) to obtain records for all domestic commercial flights in the contiguous United States over a one-week period from 7th to 13th April 2025, yielding 143,152 individual flight records. 
For each flight, the dataset provides the origin and destination airport codes, the actual departure date, wheel-off time and elapse time. 
In addition, all local timestamps are converted to coordinated universal time (UTC) to ensure consistent temporal alignment across time zones.
We use the wheel-off time as the departure timestamp and the elapsed time to compute arrival time. 
Flight trajectories are modeled as great-circle paths between origin and destination airports at a constant cruise altitude and constant cruise speed. 
This simplification is appropriate for a system-level assessment because it preserves the primary geometric relationship between flight paths and solar farm locations while avoiding the computational burden of full four-dimensional trajectory modeling across more than 140,000 flights.
Each flight trajectory is discretized into a sequence of positions at uniform time increments of $\Delta t$ to enable time-resolved evaluation of farm-flight coverage interactions.

\subsubsection{Airport data}

Airport geographic coordinates are obtained from the USA Airports dataset \cite{airportdata} maintained by Esri and distributed through ArcGIS Data and Maps. 
For each of the airports appearing in the records, we extract the latitude, longitude and Federal Aviation Administration (FAA) identifier. 
These coordinates serve as the origin and destination endpoints for flight and are used to compute great-circle distances, bearings and flight durations for each flight segment.

\subsubsection{Solar farm data}

We use the United States Large-Scale Solar Photovoltaic Database \cite{solarfarmdata} maintained by the Lawrence Berkeley National Laboratory to characterize the existing fleet of utility-scale solar installations. 
The database contains 5,712 solar farms as of the study period, each described by its name, geographic coordinates (latitude and longitude), DC capacity, ground area and site geometry (polygon boundary).
Farm geometries are visualized and validated using OpenStreetMap \cite{haklay2008openstreetmap} base layers. 
From the capacity and area fields, we derive two key quantities for each farm: the DC capacity, which determines the maximum electrical power available for conversion and transmission, and the farm aperture area, which governs the Fresnel spot size and near-field beaming geometry of the phased-array transmitter. 
Not all farms in the database are suitable for energy beaming. 
The effective transmit capacity of each farm is defined as the minimum of the safety capacity and the DC capacity. 
A farm qualifies for participation in the beaming network if and only if its effective capacity is sufficient to deliver at least the minimum useful received-power threshold $\varepsilon = 1$ MW to an aircraft at cruise altitude, after accounting for the end-to-end system efficiency and the beam geometry losses at the maximum slant range. 
Applying this criterion reduces the eligible set from 5,712 to 437 solar farms, representing 29.35 GW of aggregate safety capacity.

\subsection{System}

\subsubsection{DC power source}

Each solar farm is modeled as a DC power source that can supply electrical energy to the phased-array transmitter either directly from PV generation during daylight hours or from a co-located energy storage system and microgrid connection during nighttime and low-irradiance periods. 
We assume that each farm maintains its full capacity throughout both daytime and nighttime operating windows, reflecting the availability of grid-connected battery storage or microgrid back-up. 
The effective transmit capacity of each farm is defined as the minimum of two quantities: the safety capacity and the DC capacity:
\begin{equation}
P_{\mathrm{farm}} = \min\{P_{\mathrm{safety}},\; P_{\mathrm{dc}}\}.
\end{equation}
The safety capacity is determined by a maximum permissible ground-level power density of $S_{\mathrm{ab}} = 20$ W/m$^2$, a basic restriction for electromagnetic field exposure \cite{international2020guidelines}. 
For a farm with aperture area $A_f$, the safety capacity is:
\begin{equation}
P_{\mathrm{safety}} = S_{\mathrm{ab}} \cdot A_f.
\end{equation}
The DC capacity $P_{\mathrm{dc}}$ is the electrical capacity of the installation, directly exported from the solar farm database. 
The binding constraint between $P_{\mathrm{safety}}$ and $P_{\mathrm{dc}}$ sets the maximum power that the farm can inject into the beaming subsystem at any given time.
All solar farms in the qualified set are assumed to be equipped with the same panel type integrated with embedded phase-controlled transmitting antennas, enabling multi-directional plane tracking and precise beam synchronization with up to 90-degree maximum beam steering from the array normal.

\subsubsection{Aircraft}

All aircraft in the analysis are modeled as hybrid-electric Airbus A320, a narrow-body, single-aisle airframe that is representative of the dominant fleet type operating on domestic routes in the United States. 
The hybrid-electric configuration assumes that the aircraft is equipped with both conventional jet-fuel propulsion and an electric propulsion system powered by a receiving antenna array mounted on the fuselage. 
Key aircraft parameters include the airframe dimensions (length: 37.57 m, width: 35.8 m, height: 11.76 m, cabin width: 3.7 m) and surface area (wing area: 122.6 m$^2$, bottom surface: 139 m$^2$), a maximum cruise altitude $h_{\mathrm{cruise}} = 12,100$ m, a maximum cruise speed $v_{\mathrm{cruise}} = 235$ m/s, a cruise drag $F_{\mathrm{drag}} = 37.3$ kN, a cruise fuel flow $\dot{m}_{\mathrm{fuel}} = 2,200$ kg/h, a propulsion efficiency $\eta_{\mathrm{prop}} = 85\%$, and a maximum take-off weight of 75.5 tons \cite{airbus}. 
The cruise power demand $P_{\mathrm{cruise}}$ is:
\begin{equation}
P_{\mathrm{cruise}} = \frac{F_{\mathrm{drag}} \cdot v_{\mathrm{cruise}}}{\eta_{\mathrm{prop}}} = \frac{37.3 \times 10^3 \;\mathrm{N} \times 235 \;\mathrm{m/s}}{0.85} = 10.3 \;\mathrm{MW}.
\end{equation}
The effective receiving aperture area is $A_r = 261.6$ m$^2$, computed from the aircraft's bottom surface geometry and antenna integration constraints.
All flights are assumed to be fully loaded, consistent with a conservative assessment of fuel burn and beaming benefit. 
Flight trajectories are modeled as straight-line great-circle paths between origin and destination airports at the specified cruise altitude and constant cruise speed. 
We assume instantaneous and lossless switching between the electric and jet-fuel energy supply modes, such that the aircraft draws beamed power whenever it falls within the coverage footprint of a qualified solar farm and reverts to jet fuel otherwise. 
No onboard battery storage is assumed and beamed power is consumed directly by the electric propulsion system during the powered interval.
This binary switching assumption is conservative in the sense that it neglects potential efficiency gains from blended power modes, but it simplifies the energy accounting and avoids introducing assumptions about power management strategies that are not yet validated at the flight level.

\subsubsection{Power beaming}

The microwave power beaming subsystem converts DC electrical power at the solar farm into a focused microwave beam, transmits that beam through free space to the aircraft, and reconverts the received microwave energy into DC electrical power at the onboard rectenna. 
The end-to-end transfer efficiency $\eta_{\mathrm{sys}}$ is decomposed into four multiplicative stages:
\begin{equation}
\eta_{\mathrm{sys}} = \eta_{\mathrm{dc-rf}} \cdot \eta_{\mathrm{free}} \cdot \eta_{\mathrm{rf-dc}} \cdot \eta_{\mathrm{spot}},
\end{equation}
where $\eta_{\mathrm{dc-rf}} = 68.87\%$ is the DC-to-RF conversion efficiency at the transmitter, reflecting the performance of solid-state power amplifiers integrated into the phased-array elements \cite{greda2019beamsteering}, $\eta_{\mathrm{free}} = 95\%$ is the free-space propagation efficiency, accounting for atmospheric absorption and diffusion losses in the microwave band \cite{rodenbeck2021microwave}, $\eta_{\mathrm{rf-dc}} = 78.67\%$ is the RF-to-DC rectification efficiency at the aircraft receiver \cite{rodenbeck2021microwave}, and $\eta_{\mathrm{spot}} = 87\%$ is the spot efficiency, capturing the fraction of transmitted power that falls within the receiver aperture \cite{sheth2023concept}. 
The combined end-to-end efficiency is:
\begin{equation}
\eta_{\mathrm{sys}} = 68.87\% \times 95\% \times 78.67\% \times 87\% = 44.78\%.
\end{equation}
This value reflects current-generation component performance and is applied uniformly across all farm–aircraft links.

When electromagnetic radiation passes through an aperture, the field distribution at a distant observation plane is not a simple geometric projection of the aperture shape but is instead governed by wave interference among the secondary wavelets emanating from every point on the aperture surface. 
This interference is described by the Fresnel-Kirchhoff diffraction integral \cite{born2000principles}. 
For a planar aperture of area $A_f$ radiating a monochromatic field $U_0(x', y')$ at wavelength $\lambda$, the scalar diffracted field at a point $(x, y, z)$ is:
\begin{equation}
U(x,y,z) = \frac{1}{i\lambda z} \iint U_0(x', y')\, \exp\!\left\{\frac{i\pi\bigl[(x - x')^2 + (y - y')^2\bigr]}{\lambda z}\right\} dx'\, dy',
\end{equation}
where the integration is performed over the transmitting aperture and the expression is valid in the Fresnel approximation \cite{southwell1981validity}. 
This integral is the physical foundation for two claims made in the paper. 
First, it explains the mechanism behind the ``embedded phase-controlled transmitting antennas'': the phased-array elements apply a programmable phase distribution $U_0(x', y')$ across the aperture to shape the diffracted field at the aircraft's position. 
Second, the quadratic phase term $\exp\{i\pi[(x - x')^2 + (y - y')^2]/(\lambda z)\}$ is the mathematical object that the array manipulates to achieve beam focusing, a capability that cannot be asserted without this wave-optical basis, and that distinguishes the SoPhAr concept from conventional omnidirectional broadcast.

The diffraction integral alone does not determine whether beam focusing is physically achievable at a given distance. 
That question is answered by the Fresnel number, which characterizes the ratio of aperture-induced phase curvature to the wavelength:
\begin{equation}
N_F = \frac{D^2}{4\lambda z},
\end{equation}
where $D$ is the effective diameter of the transmitting aperture and $z$ is the slant range. 
In the radiative near-field regime ($N_F \geq 1$), the phase curvature is significant, and the array can compensate it through a conjugate phase distribution, producing constructive interference at a designated focal point and concentrating the beam energy into a small spot.
Because utility-scale solar farms have aperture dimensions on the order of hundreds of meters to kilometers, the system operates well within the radiative near-field for aircraft at typical cruise altitudes (9,000–15,000 m), enabling tight beam focusing and high spot efficiency. 
This is the quantitative basis for the claim that SoPhAr-based WPT can deliver meaningful power to aircraft at cruise altitude: the large Fresnel number guarantees that phase-controlled focusing is not merely possible but highly effective. 

In this regime, the phased-array transmitter applies a quadratic phase correction across the aperture to focus the beam at the aircraft's three-dimensional position. 
The required phase profile for focusing at slant range $z$ is:
\begin{equation}
\varphi(x', y') = -\frac{\pi(x'^2 + y'^2)}{\lambda z},
\end{equation}
which exactly compensates the quadratic phase term in the Fresnel diffraction integral, causing constructive interference of all secondary wavelets at the focal point and producing a concentrated spot of power. 
The ``precise beam synchronization with up to 90-degree maximum beam steering'' refers to the real-time computation and application of this phase profile as the aircraft moves along its trajectory, with the steering angle determined by the aircraft's lateral offset from the farm's bore-sight.

The transmitter model computes the Fresnel spot area at the aircraft altitude using the Fresnel–Kirchhoff diffraction integral, which determines the spatial distribution of the electromagnetic field in the radiative near-field region. 
For a circular transmitting aperture of effective diameter $D$ operating at wavelength $\lambda = 0.05$ m (corresponding to a microwave frequency of 6 GHz), the Fresnel spot diameter at slant range $z$ scales as the focused Airy-like pattern \cite{rivera2016wavelength}:
\begin{equation}
d_{\mathrm{spot}}(z) = \frac{2.44\,\lambda\, z}{D}.
\end{equation}
This approximation captures the central-lobe width and is sufficient for estimating the effective beam footprint.
It also justifies the modeling assumption that each farm delivers power to one aircraft at a time, since the beam is narrow enough to target individual aircraft without significant spillover to neighboring flight paths.

For a farm transmitting power $P_t$ through the beaming subsystem, the power density at slant range $z$ is:
\begin{equation}
S(z) = \frac{\eta_{\mathrm{sys}}\, P_t}{\pi\, \lambda\, z},
\end{equation}
and the received power is obtained by integrating this density over the effective area of the onboard rectenna:
\begin{equation}
P_r = S(z) \cdot A_r = \frac{\eta_{\mathrm{sys}}\, A_r}{\pi\, \lambda\, z}\; P_t,
\end{equation}
where the coefficient $\eta_{\mathrm{sys}} A_r / (\pi \lambda z)$ is the near-field power transfer coefficient that governs the fraction of transmitted power captured by the aircraft at slant range $z$.

The energy boundary of each farm defined as the ground-projected region within which the beaming system can deliver received power at or above the minimum useful threshold $\varepsilon = 1$ MW to an aircraft at cruise altitude. 
For a farm with effective capacity $P_f$, the maximum range $R_{\mathrm{beam}}$ at which the threshold is met is obtained by setting $P_r = \varepsilon$ and solving for $z$:
\begin{equation}
R_{\mathrm{beam}} = \frac{\eta_{\mathrm{sys}}\, A_r\, P_f}{\pi\, \lambda\, \varepsilon}.
\label{eq:rbeam}
\end{equation}
At the reference cruise altitude of $h_{\mathrm{cruise}} = 12,100$ m, this implies that a farm must have sufficient capacity to satisfy $R_{\mathrm{beam}} \geq h_{\mathrm{cruise}}$. 
The corresponding minimum effective capacity is:
\begin{equation}
P_f^{\min} = \frac{\pi\, \lambda\, \varepsilon\, h_{\mathrm{cruise}}}{\eta_{\mathrm{sys}}\, A_r} = \frac{\pi \times 0.05 \times 1.0 \times 12{,}100}{0.4478 \times 261.6} \approx 16.2 \;\mathrm{MW}.
\label{eq:pfmin}
\end{equation}
This qualification criterion, combined with the effective capacity definition, determines the set of 437 eligible farms. 
An aircraft is considered to be receiving beamed power at a given time step if and only if its ground-projected position falls within the energy boundary of at least one qualified farm, and the slant range $z$ between the aircraft and that farm satisfies $z \leq R_{\mathrm{beam}}$.

\subsection{Optimization}

We formulate two complementary optimization problems to quantify the maximum achievable system-wide benefits of SoPhAr-enabled energy beaming under realistic operational constraints. 
The first, flight schedule optimization, determines the optimal departure time adjustments for all flights given a fixed set of participating solar farms and aircraft. 
The second, farm-and-flight choice optimization, selects the subset of solar farms to equip with SoPhAr functionality and the subset of flights to equip with receiving antennas, given exogenous market penetration rates. 
Both problems share a common objective structure and a common set of system parameters, but differ in their decision variables and constraint sets. 
We first introduce the notation common to both formulations before presenting each problem separately.

Let $I$ denote the set of flights, $F$ the set of solar farms, $T$ the set of discrete time steps, and $S_i = \{-\tau_{\max}, -\tau_{\max} + \Delta t, \ldots, \tau_{\max}\}$ the set of admissible time-shift options for flight $i \in I$, where $\tau_{\max} = 1{,}800$~s is the maximum allowable schedule adjustment and $\Delta t$ is the time-step resolution. 
The system parameters governing the DC power source, the aircraft, the power beaming subsystem, and time and coverage are listed in supplementary materials and summarized below for convenience.

For the aircraft, we denote the cruise fuel burn rate by $\dot{m}_{\mathrm{fuel}}$ (kg/h), the unit fuel price by $C_{\mathrm{fuel}}$ (\$/kg), the CO2 emission factor per unit fuel by $e_{\mathrm{fuel}}$ (kg CO2/kg fuel), the cruise power demand by $P_{\mathrm{cruise}}$ (MW), the cruise speed by $v_{\mathrm{cruise}}$ (m/s), and the receiving aperture area by $A_r$ (m$^2$). 
For each solar farm $f \in F$, $P_f$ denotes its effective capacity (MW) and $A_f$ the farm aperture area (m$^2$). 
The end-to-end system efficiency is $\eta_{\mathrm{sys}}$, and the solar electricity cost is $C_{\mathrm{elec}}$ (\$/MWh) with associated emission intensity $e_{\mathrm{solar}}$ (kg CO2/MWh). 
We denote the social cost of carbon by $C_{\mathrm{CO2}}$ (\$/kg CO2), which monetizes the climate externality of both jet fuel combustion and solar electricity generation.
The operating wavelength is $\lambda$ (m), and $z_{f,i,t}$ (or $z_{f,i,t,s}$ when schedule shifting is active) denotes the slant-range distance between farm $f$ and flight $i$ at time $t$.

We define the following decision variables common to both formulations. 
For each flight $i \in I$ and time step $t \in T$, let $b_{i,t} \in \{0,1\}$ be a binary indicator equal to 1 if flight $i$ receives beamed power at time~$t$; let $r_{i,t} \geq 0$ be the total received power at flight $i$ at time~$t$; and for each farm-flight-time triple, let $p_{f,i,t} \geq 0$ (or $p_{f,i,t,s}$ in the flight schedule optimization) be the power transmitted from farm $f$ to flight $i$ at time $t$. 
The binary coverage parameter $a_{f,i,t} \in \{0,1\}$ (or $a_{f,i,t,s}$) indicates whether flight $i$ lies within the energy boundary of farm $f$ at time $t$, as determined by the beam geometry model. 
A minimum received-power threshold $\varepsilon = 1$ MW defines the lower bound for meaningful power transfer.

\subsubsection{Objective function}

Both optimization problems share a common objective that maximizes the net economic benefit of replacing jet fuel with beamed solar electricity. 
The benefit comprises two components: the avoided cost of jet fuel consumption (including both the direct fuel expenditure and the associated CO2 cost) minus the cost of producing and transmitting the replacement solar electricity. 
Specifically, the objective function is:
\begin{equation}
\begin{aligned}
\max \quad Z &= \underbrace{\left(\dot{m}_{\mathrm{fuel}} \cdot C_{\mathrm{fuel}} + \dot{m}_{\mathrm{fuel}} \cdot e_{\mathrm{fuel}} \cdot C_{\mathrm{CO2}} \right) \cdot \frac{1}{P_{\mathrm{cruise}}}}_{\text{fuel and CO2}\text{ saving rate}} \; \sum_{i \in I} \sum_{t \in T} \Delta t \cdot r_{i,t} \; \\ &-\; \underbrace{\left(C_{\mathrm{elec}} + e_{\mathrm{solar}} \cdot C_{\mathrm{CO2}}\right)}_{\text{electricity and CO2}\text{ cost rate}} \; \sum_{i \in I} \sum_{t \in T} \Delta t \cdot r_{i,t},
\end{aligned}
\end{equation}
where the first term captures the monetary value of jet fuel and carbon emissions avoided per unit of beamed energy received, and the second term accounts for the cost of generating and delivering that energy via the SoPhAr system. 
The three objectives: maximizing fuel saving, minimizing carbon emissions, and maximizing beam utilization, are thereby captured in a single scalar objective that reflects their economic equivalence under the assumed cost parameters.
The total cost reduction decomposes additively into fuel cost saving (the $C_{\mathrm{fuel}}$ component of the first term minus the $C_{\mathrm{elec}}$ component of the second) and CO2 cost saving (the $e_{\mathrm{fuel}} \cdot C_{\mathrm{CO2}}$ component minus the $e_{\mathrm{solar}} \cdot C_{\mathrm{CO2}}$ component), enabling separate tracking of the economic and environmental value streams reported in the results.

\subsubsection{Flight schedule optimization}

The flight schedule optimization problem determines, for each flight $i \in I$, the optimal departure time shift $s \in S_i$ that maximizes system-wide net benefit, assuming that all solar farms and all flights participate in the beaming network (i.e., 100\% penetration of both farms and flights). 
In addition to the common variables, we introduce two sets of binary decision variables: $d_{i,s} \in \{0,1\}$, which equals 1 if flight $i$ is shifted by $s$ time steps, and $c_{i,t,s} \in \{0,1\}$, which equals 1 if flight $i$ is airborne at global time $t$ under schedule shift $s$.
The flight schedule optimization problem is formulated as:
\begin{equation}
\max_{d, p, r, b} \quad Z(r),
\end{equation}
subject to the following constraints.

\paragraph{Time-shift constraint} 
Each flight must be assigned exactly one departure time shift:
\begin{equation}
\sum_{s \in S_i} d_{i,s} = 1, \qquad \forall\, i \in I.
\label{eq:time_shift}
\end{equation}

\paragraph{Capacity constraints} 
The power transmitted from any farm to any flight must respect the farm's effective capacity and can only be non-zero when the flight is within beaming range under the selected shift:
\begin{align}
0 \leq p_{f,i,t,s} &\leq P_f \cdot a_{f,i,t,s} \cdot d_{i,s}, &\forall\, f \in F,\; i \in I,\; t \in T,\; s \in S_i, \\
\sum_{i \in I} \sum_{s \in S_i} p_{f,i,t,s} &\leq P_f, &\forall\, f \in F,\; t \in T,
\end{align}
where Eq. (17) ensures that power is allocated to a flight only if that flight falls within the energy boundary of the farm under the chosen time shift, i.e., $a_{f,i,t,s} = 1$ and $d_{i,s} = 1$, and Eq. (18) limits the aggregate power dispatched by each farm at any time step to its effective capacity.

\paragraph{Received power constraints} 
The total power received by each flight is determined by the beaming physics and the farm-flight geometry:
\begin{align}
r_{i,t} &= \sum_{f \in F} \sum_{s \in S_i} \frac{\eta_{\mathrm{sys}}\, A_r}{\pi\, \lambda\, z_{f,i,t,s}}\; p_{f,i,t,s}, &\forall\, i \in I,\; t \in T, \\
0 \leq r_{i,t} &\leq P_f \cdot \sum_{s \in S_i} c_{i,t,s}\, d_{i,s}, &\forall\, i \in I,\; t \in T,
\end{align}
where Eq. (19) computes the received power as the sum of contributions from all farms, weighted by the near-field power transfer coefficient $\eta_{\mathrm{sys}} A_r / (\pi \lambda z_{f,i,t,s})$, which depends on the system efficiency, the receiver aperture area, the wavelength, and the slant range, and Eq. (20) enforces that power can only be received when the flight is airborne under its selected schedule shift.

\paragraph{Flight participation constraints} 
These constraints link the beaming indicator to the flight status and establish the minimum useful power threshold:
\begin{align}
b_{i,t} &\leq \sum_{s \in S_i} c_{i,t,s}\, d_{i,s}, &\forall\, i \in I,\; t \in T, \\
r_{i,t} &\geq \varepsilon\, b_{i,t}, &\forall\, i \in I,\; t \in T, \\
r_{i,t} &\leq P_f\, b_{i,t}, &\forall\, i \in I,\; t \in T,
\end{align}
where Eq. (21) ensures that a flight can only be marked as receiving power if it is currently in flight, Eq. (22) establishes a minimum received-power threshold $\varepsilon$, so that negligible power transfers are not counted as active beaming events, and Eq. (23) provides the complementary upper bound, ensuring that $b_{i,t} = 0$ forces $r_{i,t} = 0$.

This problem is evaluated at three cruise altitudes: 9,100 m, 12,100 m and 15,100 m to examine the interaction between altitude and flight schedule coordination. 
Higher altitudes expand the effective beaming footprint, increasing both the number of farm-flight interactions and the potential for flight schedule optimization to align flights with available beaming windows.

\subsubsection{Farm-and-flight choice optimization}
\label{sec:ffco}

The farm-and-flight choice optimization problem determines which solar farms should be equipped with SoPhAr system and which flights should be equipped with receiving antennas, given exogenous penetration rates $\rho_F$ for farms and $\rho_I$ for flights. 
Unlike the flight schedule optimization, this formulation does not permit time shifting of flight schedules; instead, it selects the most valuable subset of solar farm and flight to participate in the beaming network under fixed timetables. 
The additional binary decision variables are: $x_f \in \{0,1\}$, which equals 1 if solar farm $f$ is equipped with SoPhAr system, and $x_i \in \{0,1\}$, which equals 1 if flight $i$ is equipped with a receiving antenna.
The farm-and-flight choice optimization problem is formulated as:
\begin{equation}
\max_{x_f, x_i, p, r, b} \quad Z(r),
\end{equation}
subject to the following constraints.

\paragraph{Capacity constraints} 
Power can only be transmitted from an equipped farm to a flight within its beaming range:
\begin{align}
0 \leq p_{f,i,t} &\leq P_f \cdot x_f \cdot a_{f,i,t}, &\forall\, f \in F,\; i \in I,\; t \in T, \\
\sum_{i \in I} p_{f,i,t} &\leq P_f \cdot x_f, &\forall\, f \in F,\; t \in T,
\end{align}
where Eq. (25) couples the power allocation to the farm selection variable $x_f$ and the coverage indicator $a_{f,i,t}$, and Eq. (26) limits total power dispatch to the farm's capacity, conditional on the farm being selected.

\paragraph{Penetration constraints} 
The number of equipped farms and flights must match the prescribed market penetration rates:
\begin{align}
\sum_{f \in F} x_f &= \rho_F \cdot |F|, \\
\sum_{i \in I} x_i &= \rho_I \cdot |I|,
\end{align}
where the constraints parameterize the optimization across a range of adoption scenarios. 
In the analysis, we evaluate all combinations of $\rho_F, \rho_I \in \{10\%, 20\%, \ldots, 100\%\}$, yielding 100 scenarios that span the full penetration landscape and reveal how system-level benefits scale with solar farm and flight adoption.

\paragraph{Received power constraints} 
The received power at each flight reflects the beaming physics under fixed flight schedules:
\begin{align}
r_{i,t} &= \sum_{f \in F} \frac{\eta_{\mathrm{sys}}\, A_r}{\pi\, \lambda\, z_{f,i,t}}\; p_{f,i,t}, &\forall\, i \in I,\; t \in T, \\
0 \leq r_{i,t} &\leq P_f \cdot x_i, &\forall\, i \in I,\; t \in T,
\end{align}
where Eq. (29) computes the total received power from all selected farms, and Eq. (30) ensures that a flight can receive power only if it is equipped with a receiving antenna.

\paragraph{Flight participation constraints} 
These constraints parallel those of the schedule optimization but link the beaming indicator to the flight selection variable:
\begin{align}
b_{i,t} &\leq x_i, &\forall\, i \in I,\; t \in T, \\
r_{i,t} &\geq \varepsilon\, b_{i,t}, &\forall\, i \in I,\; t \in T, \\
r_{i,t} &\leq P_f\, b_{i,t}, &\forall\, i \in I,\; t \in T,
\end{align}
where Eq. (31) restricts the beaming indicator to flights that have been selected for antenna installation, and Eq. (32) and Eq. (33) enforce the same minimum-threshold and complementary upper bound as in the flight schedule optimization.

\subsubsection{Relationship between the two optimization formulations}

The two optimization problems address complementary planning questions and are designed to be applied at different stages of system analysis. 
The flight schedule optimization quantifies the additional value that modest timetable coordination can unlock when the full solar farm and flight are already committed, and is evaluated at three cruise altitudes (9,100 m, 12,100 m, and 15,100 m) to examine the interaction between altitude and flight scheduling. 
The farm-and-flight choice optimization addresses the prior infrastructure planning question: given that only a fraction of solar farms and flights will participate in any realistic near-term deployment, which subset should be prioritized? 
Together, these two optimization formulations reveal both the intensive margin (how much more value can be extracted per participant through coordination) and the extensive margin (how value scales with the number of participants), providing a comprehensive characterization of the economic structure of SoPhAr-based WPT for commercial aviation.

Both problems are formulated as MILPs. 
The coverage parameters $a_{f,i,t}$ and $a_{f,i,t,s}$ and the in-flight indicators $c_{i,t,s}$ are pre-computed from the flight trajectory and solar farm geometry data, and are treated as fixed binary inputs to the optimization. 
This pre-computation substantially reduces the online computational burden, as the geometric relationships between all farm-flight pairs across all time steps are resolved prior to optimization. 
The resulting MILPs are solved using a commercial solver Gurobi \cite{kemminer2024configuring}.

\backmatter

\bmhead{Supplementary information}

The supplementary materials are available.

\bmhead{Acknowledgements}

This research was supported by the NSF Grants (2125858, 2236305), UT Good Systems Grand Challenge, the MITRE Corporation, and UT Energy Institute Strategic Energy Seed Project.

\section*{Declarations}

\begin{itemize}
\item Competing interests: The authors declare no competing interests.
\item Data availability: We use only publicly available data, with relevant references and code available. Source data necessary to replicate the results in the paper are provided via Github after acceptance.
\item Code availability: The code for this study, including analysis and visualization, can be accessed via Github after acceptance.
\item Author contribution: Conceptualization: T.W., J.M., K.K. and C.C. Data curation: T.W., Y.X. and J.B. Formal analysis: T.W., Y.X. and J.B. Funding acquisition: J.J., J.M., K.K. and C.C. Investigation: T.W., Y.X., J.B., J.J., J.M., K.K. and C.C. Methodology: T.W., Y.X., J.B., J.M., K.K. and C.C. Project administration: C.C. Resources: J.J., J.M., K.K. and C.C. Software: T.W. and Y.X. Supervision: J.J., J.M., K.K., C.C. and A.B. Validation: T.W. and Y.X. Visualization: T.W., Y.X. and J.B. Writing–original draft: T.W., Y.X., J.B. and C.C. Writing–review and editing: T.W., Y.X., J.B., J.J., J.M., K.K., C.C. and A.B.
\end{itemize}


\bibliography{sn-bibliography}

@article{allan2023intergovernmental,
  title={Intergovernmental panel on climate change (IPCC). Summary for policymakers},
  author={Allan, Richard P and Arias, Paola A and Berger, Sophie and Canadell, Josep G and Cassou, Christophe and Chen, Deliang and Cherchi, Annalisa and Connors, Sarah L and Coppola, Erika and Cruz, Faye Abigail and others},
  journal={Climate Change 2021: The Physical Science Basis. Contribution of Working Group I to the Sixth Assessment Report of the Intergovernmental Panel on Climate Change},
  pages={3--32},
  year={2023},
  publisher={Cambridge University Press}
}

@article{dray2022cost,
  title={Cost and emissions pathways towards net-zero climate impacts in aviation},
  author={Dray, Lynnette and Sch{\"a}fer, Andreas W and Grobler, Carla and Falter, Christoph and Allroggen, Florian and Stettler, Marc EJ and Barrett, Steven RH},
  journal={Nature Climate Change},
  volume={12},
  number={10},
  pages={956--962},
  year={2022},
  publisher={Nature Publishing Group UK London}
}

@article{bergero2023pathways,
  title={Pathways to net-zero emissions from aviation},
  author={Bergero, Candelaria and Gosnell, Greer and Gielen, Dolf and Kang, Seungwoo and Bazilian, Morgan and Davis, Steven J},
  journal={Nature Sustainability},
  volume={6},
  number={4},
  pages={404--414},
  year={2023},
  publisher={Nature Publishing Group UK London}
}

@inproceedings{kosir2019improvement,
  title={Improvement in jet aircraft operation with the use of high-performance drop-in fuels},
  author={Kosir, Shane T and Behnke, Lily and Heyne, Joshua S and Stachler, Robert D and Flora, Giacomo and Zabarnick, Steven and George, Anthe and Landera, Alexander and Bambha, Ray and Denney, Russell and others},
  booktitle={AIAA Scitech Forum},
  pages={0993},
  year={2019}
}

@article{male2021us,
  title={The US energy system and the production of sustainable aviation fuel from clean electricity},
  author={Male, Jonathan L and Kintner-Meyer, Michael CW and Weber, Robert S},
  journal={Frontiers in Energy Research},
  volume={9},
  pages={765360},
  year={2021},
  publisher={Frontiers Media SA}
}

@article{vardon2022realizing,
  title={Realizing “net-zero-carbon” sustainable aviation fuel},
  author={Vardon, Derek R and Sherbacow, Bryan J and Guan, Kaiyu and Heyne, Joshua S and Abdullah, Zia},
  journal={Joule},
  volume={6},
  number={1},
  pages={16--21},
  year={2022},
  publisher={Elsevier}
}

@article{watson2024sustainable,
  title={Sustainable aviation fuel technologies, costs, emissions, policies, and markets: A critical review},
  author={Watson, Madelynn J and Machado, Pedro Gerber and Da Silva, AV and Saltar, Y and Ribeiro, CO and Nascimento, Cl{\'a}udio Augusto Oller do and Dowling, Alexander W},
  journal={Journal of Cleaner Production},
  volume={449},
  pages={141472},
  year={2024},
  publisher={Elsevier}
}

@article{yusaf2024sustainable,
  title={Sustainable hydrogen energy in aviation--A narrative review},
  author={Yusaf, Talal and Mahamude, Abu Shadate Faisal and Kadirgama, Kumaran and Ramasamy, Devarajan and Farhana, Kaniz and Dhahad, Hayder A and Talib, ABD Rahim Abu},
  journal={International Journal of Hydrogen Energy},
  volume={52},
  pages={1026--1045},
  year={2024},
  publisher={Elsevier}
}

@article{viswanathan2019potential,
  title={Potential for electric aircraft},
  author={Viswanathan, Venkatasubramanian and Knapp, B Matthew},
  journal={Nature Sustainability},
  volume={2},
  number={2},
  pages={88--89},
  year={2019},
  publisher={Nature Publishing Group UK London}
}

@article{schafer2019technological,
  title={Technological, economic and environmental prospects of all-electric aircraft},
  author={Sch{\"a}fer, Andreas W and Barrett, Steven RH and Doyme, Khan and Dray, Lynnette M and Gnadt, Albert R and Self, Rod and O’Sullivan, Aidan and Synodinos, Athanasios P and Torija, Antonio J},
  journal={Nature Energy},
  volume={4},
  number={2},
  pages={160--166},
  year={2019},
  publisher={Nature Publishing Group UK London}
}

@inproceedings{sampson2024operation,
  title={Operation and Performance Optimization of an All-Electric, Narrow-Body Airliner},
  author={Sampson, Ariel D and Drake, Aaron},
  booktitle={AIAA AVIATION FORUM AND ASCEND},
  pages={4185},
  year={2024}
}

@article{wheeler2021electric,
  title={Electric/hybrid-electric aircraft propulsion systems},
  author={Wheeler, Patrick and Sirimanna, Thusara Samith and Bozhko, Serhiy and Haran, Kiruba S},
  journal={Proceedings of the IEEE},
  volume={109},
  number={6},
  pages={1115--1127},
  year={2021},
  publisher={IEEE}
}

@article{rendon2021aircraft,
  title={Aircraft hybrid-electric propulsion: Development trends, challenges and opportunities},
  author={Rend{\'o}n, Manuel A and S{\'a}nchez R, Carlos D and Gallo M, Josselyn and Anzai, Alexandre H},
  journal={Journal of Control, Automation and Electrical Systems},
  volume={32},
  number={5},
  pages={1244--1268},
  year={2021},
  publisher={Springer}
}

@article{sayed2021review,
  title={Review of electric machines in more-/hybrid-/turbo-electric aircraft},
  author={Sayed, Ehab and Abdalmagid, Mohamed and Pietrini, Giorgio and Sa’adeh, Nicole-Marie and Callegaro, Alan Dorneles and Goldstein, Cyrille and Emadi, Ali},
  journal={IEEE Transactions on Transportation Electrification},
  volume={7},
  number={4},
  pages={2976--3005},
  year={2021},
  publisher={IEEE}
}

@article{viswanathan2022challenges,
  title={The challenges and opportunities of battery-powered flight},
  author={Viswanathan, Venkatasubramanian and Epstein, Alan H and Chiang, Yet-Ming and Takeuchi, Esther and Bradley, Marty and Langford, John and Winter, Michael},
  journal={Nature},
  volume={601},
  number={7894},
  pages={519--525},
  year={2022},
  publisher={Nature Publishing Group UK London}
}

@article{mehedi2022life,
  title={Life cycle greenhouse gas emissions and energy footprints of utility-scale solar energy systems},
  author={Mehedi, Tanveer Hassan and Gemechu, Eskinder and Kumar, Amit},
  journal={Applied Energy},
  volume={314},
  pages={118918},
  year={2022},
  publisher={Elsevier}
}

@article{aleksandra2024role,
  title={Role of solar PV in net-zero growth: An analysis of international manufacturers and policies},
  author={Aleksandra, Arcipowska and Sara, Blanco Perez and Ma{\l}gorzata, Jakim{\'o}w and Brian, Baldassarre and Davide, Polverini and Miguel, Cabrera},
  journal={Progress in Photovoltaics: Research and Applications},
  volume={32},
  number={9},
  pages={607--622},
  year={2024},
  publisher={Wiley Online Library}
}

@techreport{prasanna2021storage,
  title={Storage futures study: Distributed solar and storage outlook: Methodology and scenarios},
  author={Prasanna, Ashreeta and McCabe, Kevin and Sigrin, Ben and Blair, Nathan},
  year={2021},
  institution={National Renewable Energy Lab.(NREL), Golden, CO (United States)}
}

@techreport{heath2022environmental,
  title={Environmental and circular economy implications of solar energy in a decarbonized US grid},
  author={Heath, Garvin and Ravikumar, Dwarakanath and Ovaitt, Silvana and Walston, Leroy and Curtis, Taylor and Millstein, Dev and Mirletz, Heather and Hartmann, Heidi and McCall, James},
  year={2022},
  institution={National Renewable Energy Laboratory (NREL), Golden, CO (United States)}
}

@article{brunet2022will,
  title={Will solar energy escape the natural “resource curse”?},
  author={Brunet, Carole and Bouchard, Michel A and Baptiste, Pierre and Savadogo, Oumarou and Sokona, Youba and Merveille, Nicolas},
  journal={Energy Strategy Reviews},
  volume={44},
  pages={101010},
  year={2022},
  publisher={Elsevier}
}

@article{maka2024pathway,
  title={The pathway towards decarbonisation and net-zero emissions by 2050: The role of solar energy technology},
  author={Maka, Ali OM and Ghalut, Tarik and Elsaye, Elsaye},
  journal={Green Technologies and Sustainability},
  volume={2},
  number={3},
  pages={100107},
  year={2024},
  publisher={Elsevier}
}

@article{rodenbeck2021microwave,
  title={Microwave and millimeter wave power beaming},
  author={Rodenbeck, Christopher T and Jaffe, Paul I and Strassner II, Bernd H and Hausgen, Paul E and McSpadden, James O and Kazemi, Hooman and Shinohara, Naoki and Tierney, Brian B and DePuma, Christopher B and Self, Amanda P},
  journal={IEEE Journal of Microwaves},
  volume={1},
  number={1},
  pages={229--259},
  year={2021},
  publisher={IEEE}
}

@article{shinohara2013beam,
  title={Beam control technologies with a high-efficiency phased array for microwave power transmission in Japan},
  author={Shinohara, Naoki},
  journal={Proceedings of the IEEE},
  volume={101},
  number={6},
  pages={1448--1463},
  year={2013},
  publisher={IEEE}
}

@inproceedings{orndorff2023gradient,
  title={Gradient-based sizing optimization of power-beaming-enabled aircraft},
  author={Orndorff, Nicholas C and Wang, Bingran and Ruh, Marius L and Fletcher, Andrew and Hwang, John T},
  booktitle={AIAA Aviation Forum},
  pages={4019},
  year={2023}
}

@inproceedings{wang2025large,
  title={Large-scale MDO under uncertainty of an eVTOL aircraft using dimension reduction via global sensitivity analysis},
  author={Wang, Bingran and Ruh, Marius L and Tian, Aoran and Scotzniovsky, Luca and Hwang, John T},
  booktitle={AIAA AVIATION FORUM AND ASCEND},
  pages={3344},
  year={2025}
}

@article{brown2007experiments,
  title={Experiments involving a microwave beam to power and position a helicopter},
  author={Brown, William C},
  journal={IEEE Transactions on Aerospace and Electronic Systems},
  volume={AES-5},
  number={5},
  pages={692--702},
  year={2007},
  publisher={IEEE}
}

@article{mahbub2024far,
  title={Far-field wireless power beaming to mobile receivers using distributed, coherent phased arrays: A review of the critical components of a distributed wireless power beaming system},
  author={Mahbub, Ifana and Patwary, Adnan Basir and Mahin, Rafsan and Roy, Sunanda},
  journal={IEEE Microwave Magazine},
  volume={25},
  number={5},
  pages={72--94},
  year={2024},
  publisher={IEEE}
}

@inproceedings{zhao2020ka,
  title={Ka-band based channel modeling and analysis in high altitude platform (HAP) system},
  author={Zhao, Jiarui and Wang, Qi and Li, Yitao and Zhou, Jiaxi and Zhou, Wuyang},
  booktitle={IEEE Vehicular Technology Conference},
  pages={1--5},
  year={2020},
  organization={IEEE}
}

@article{yu2018design,
  title={Design of near-field focused metasurface for high-efficient wireless power transfer with multifocus characteristics},
  author={Yu, Shixing and Liu, Haixia and Li, Long},
  journal={IEEE Transactions on Industrial Electronics},
  volume={66},
  number={5},
  pages={3993--4002},
  year={2018},
  publisher={IEEE}
}

@article{claudel2024sophar,
  title={SoPhAr: Solar Phased-Arrays to boost the range of electric, hydrogen and SAF airliners in a solar world},
  author={Claudel, Christian},
  journal={arXiv preprint arXiv:2404.04779},
  year={2024}
}

@article{silva2025multibeam,
  title={Multibeam Beamforming Technology in Microwave Power Transfer and Harvesting},
  author={Silva, F{\'a}bio and Pinho, Pedro and Carvalho, Nuno Borges},
  journal={IEEE Journal of Microwaves},
  year={2025},
  publisher={IEEE}
}

@article{liu2024multifunctional,
  title={Multifunctional reconfigurable intelligent surface for wideband beamforming and frequency-and-spatial-diverse microwave sensing},
  author={Liu, Baiyang and Zhang, Qingfeng and Wong, Hang},
  journal={IEEE Transactions on Antennas and Propagation},
  year={2024},
  publisher={IEEE}
}

@article{saha2024novel,
  title={A novel theoretical modeling of the received power for phased array-based wireless power transfer system in the near-field region},
  author={Saha, Nabanita and Alvarez, Erik Pineda and Mahbub, Ifana},
  journal={IEEE Open Journal of Antennas and Propagation},
  volume={5},
  number={6},
  pages={1476--1488},
  year={2024},
  publisher={IEEE}
}

@article{patwary20244,
  title={4$\times$ 4 UWB phased array antenna with> 51 far-field scanning range for wireless power transfer application},
  author={Patwary, Adnan Basir and Mahbub, Ifana},
  journal={IEEE Open Journal of Antennas and Propagation},
  volume={5},
  number={2},
  pages={354--367},
  year={2024},
  publisher={IEEE}
}

@article{xia2024adaptive,
  title={Adaptive wireless-powered network based on CNN near-field positioning by a dual-band metasurface},
  author={Xia, De Xiao and Han, Jia Qi and Mu, Ya Jie and Guan, Lei and Wang, Xin and Ma, Xiang Jin and Zhu, Li Hao and Lv, Tian Guang and Liu, Hai Xia and Shi, Yan and others},
  journal={Nature Communications},
  volume={15},
  number={1},
  pages={10358},
  year={2024},
  publisher={Nature Publishing Group UK London}
}

@article{roy2025wireless,
  title={A wireless power beaming system using Y shaped reflectarray and rectenna achieving 8.33\% power transfer efficiency},
  author={Roy, Sunanda and Saha, Nabanita and Mahbub, Ifana},
  journal={Scientific Reports},
  volume={15},
  number={1},
  pages={17677},
  year={2025},
  publisher={Nature Publishing Group UK London}
}

@misc{flightdata,
author = {TranStats},
title = {Reporting Carrier On-Time Performance},
year = {2025},
url = {https://transtats.bts.gov/DL_SelectFields.aspx?gnoyr_VQ=FGJ&QO_fu146_anzr=b0-gvzr}
}

@misc{airportdata,
author = {Esri},
title = {USA Airports},
year = {2025},
url = {https://hub.arcgis.com/maps/5d93352406744d658d9c1f43f12b560c/about}
}

@misc{solarfarmdata,
author = {Fujita, K Sydny and Kramer, Louisa and Wellman, Kari and Diffendorfer, James and Robson, Dana and Hoen, Ben and Garrity, Chris},
title = {US Large-Scale Solar Photovoltaic Database},
year = {2025},
url = {https://emp.lbl.gov/publications/us-large-scale-solar-photovoltaic}
}

@article{haklay2008openstreetmap,
  title={Openstreetmap: User-generated street maps},
  author={Haklay, Mordechai and Weber, Patrick},
  journal={IEEE Pervasive Computing},
  volume={7},
  number={4},
  pages={12--18},
  year={2008},
  publisher={IEEE}
}

@article{international2020guidelines,
  title={Guidelines for limiting exposure to electromagnetic fields (100 kHz to 300 GHz)},
  author={International Commission on Non-Ionizing Radiation Protection and others},
  journal={Health Physics},
  volume={118},
  number={5},
  pages={483--524},
  year={2020}
}

@misc{airbus,
author = {Airbus},
title = {A320neo the benchmark},
year = {2025},
url = {https://www.aircraft.airbus.com/en/aircraft/a320-family/a320neo}
}

@article{greda2019beamsteering,
  title={Beamsteering and beamshaping using a linear antenna array based on particle swarm optimization},
  author={Greda, Lukasz A and Winterstein, Andreas and Lemes, Daniel L and Heckler, Marcos VT},
  journal={IEEE Access},
  volume={7},
  pages={141562--141573},
  year={2019},
  publisher={IEEE}
}

@inproceedings{sheth2023concept,
  title={A Concept of Operations for Power Beaming of Electric Air Vehicles},
  author={Sheth, Kapil and Schisler, Seth and Stinchfield, Todd and Winsor, Robert and Klopp, Hayden and Landis, Geoffrey and Kohlman, Lee and Pike, David},
  booktitle={IEEE/AIAA Digital Avionics Systems Conference},
  pages={1--8},
  year={2023},
  organization={IEEE}
}

@inproceedings{kemminer2024configuring,
  title={Configuring mixed-integer programming solvers for large-scale instances},
  author={Kemminer, Robin and Lange, Jannick and Kempkes, Jens Peter and Tierney, Kevin and Wei{\ss}, Dimitri},
  booktitle={Operations Research Forum},
  volume={5},
  number={2},
  pages={48},
  year={2024},
  organization={Springer}
}

@article{born2000principles,
  title={Principles of Optics: Electromagnetic Theory of Propagation, Interference and Diffraction of Light},
  author={Born, Max and Wolf, Emil and Hecht, Eugene},
  journal={Physics Today},
  volume={53},
  number={10},
  pages={77--78},
  year={2000},
  publisher={American Institute of Physics}
}

@article{southwell1981validity,
  title={Validity of the Fresnel approximation in the near field},
  author={Southwell, WH},
  journal={Journal of the Optical Society of America},
  volume={71},
  number={1},
  pages={7--14},
  year={1981},
  publisher={Optical Society of America}
}

@article{rivera2016wavelength,
  title={Wavelength estimation by using the Airy disk from a diffraction pattern with didactic purposes},
  author={Rivera-Ortega, Uriel and Pico-Gonzalez, Beatriz},
  journal={Physics Education},
  volume={51},
  number={1},
  pages={015012},
  year={2016},
  publisher={IOP Publishing}
}

\end{document}